\documentclass[
  journal=pasa,
  manuscript=research-paper, 
  year=2023,
  volume=40,
]{cup-journal}

\usepackage{microtype,siunitx,booktabs}
\usepackage{graphicx}
\usepackage{amsmath}
\usepackage{amssymb}	
\usepackage{multirow}   
\usepackage{makecell}   
\usepackage{threeparttable}  
\usepackage{booktabs}   
\usepackage{natbib}
\usepackage{xcolor}

\newcommand{\lya}{Ly$\alpha$}
\newcommand{\lymana}{Lyman-$\alpha$}

\newcommand{\HI}{H\textsc{I}}

\newcommand{\UGR}{$U_nG\cal{R}$}
\newcommand{\ugri}{$ugri$}

\title[\lya\ Spectral Type Selection of $z\sim2-3$ LBGs] {\lymana\ at Cosmic Noon I: \lya\ Spectral Type Selection of $z\sim2-3$ Lyman Break Galaxies with Broadband Imaging}

\author{Garry Foran}
\affiliation{Centre for Astrophysics and Supercomputing, Swinburne University of Technology, PO Box 218, H29, Hawthorn, VIC 3122, Australia}
\alsoaffiliation{Australian Research Council Centre of Excellence for All-sky Astrophysics in 3 Dimensions (ASTRO-3D)}
\email{gforan@swin.edu.au}

\author{Jeff Cooke}
\affiliation{Centre for Astrophysics and Supercomputing, Swinburne University of Technology, PO Box 218, H29, Hawthorn, VIC 3122, Australia}
\alsoaffiliation{Australian Research Council Centre of Excellence for All-sky Astrophysics in 3 Dimensions (ASTRO-3D)}

\author{Naveen Reddy}
\affiliation{Department of Physics \& Astronomy, University of California, Riverside, 900 University Avenue, Riverside, CA 92521, USA}

\author{Charles Steidel}
\affiliation{Cahill Center for Astronomy and Astrophysics, California Institute of Technology, MS 249-17, Pasadena, CA, 92115, USA}

\author{Alice Shapley}
\affiliation{Department of Physics \& Astronomy, University of California, Los Angeles, CA 90095, USA}

\doi{10.1017/pasa.2023.48}
\received {dd Mmm YYYY}
\revised  {dd Mmm YYYY}
\accepted {dd Mmm YYYY}
\published{}


\begin{document}

\begin{abstract} 
High-redshift Lyman break galaxies (LBGs) are efficiently selected in deep images using as few as three broadband filters, and have been shown to have multiple intrinsic and small- to large-scale environmental properties related to \lymana.  In this paper we demonstrate a statistical relationship between net \lymana\ equivalent width (net \lya\ EW) and the optical broadband photometric properties of LBGs at $z\sim2$.  We show that LBGs with the strongest net \lya\ EW in absorption (aLBGs) and strongest net \lya\ EW in emission (eLBGs) separate into overlapping but discrete distributions in $(U_n-\cal{R})$ colour and $\cal{R}$-band magnitude space, and use this segregation behaviour to determine photometric selection criteria by which sub-samples with a desired \lya\ spectral type can be selected using data from as few as three broadband optical filters.  We propose application of our result to current and future large-area and all-sky photometric surveys that will select hundreds of millions of LBGs across many hundreds to thousands of Mpc, and for which spectroscopic follow-up to obtain \lya\ spectral information is prohibitive.  To this end, we use spectrophotometry of composite spectra derived from a sample of 798 LBGs divided into quartiles on the basis of net \lya\ EW to calculate selection criteria for the isolation of \lya-absorbing and \lya-emitting populations of $z\sim3$ LBGs using \ugri\ broadband photometric data from the Vera Rubin Observatory Legacy Survey of Space and Time (LSST).
\end{abstract}

\section{INTRODUCTION }
\label{sec:intro}

One of the most important and well-studied populations of early star-forming galaxies (SFGs) are the so-called Lyman break galaxies (LBGs) that can be selected based on their rest-frame ultraviolet (UV) colours using as few as three broadband optical filters. The generic Lyman break selection method uses broadband optical photometry sensitive to the discontinuity (`break' or 'drop-out') in the rest-frame UV spectrum of SFGs blueward of the Lyman limit (912 \AA), the decrement in flux in the \lymana\ forest blueward of the \lymana\ spectral feature (\lya, 1216 \AA), and the relatively flat rest-frame UV continuum redward of \lya\ to efficiently select LBGs in large numbers, on large scales and across a wide range of redshift pathlengths.

A notable strength of the Lyman break technique is its ability to isolate populations of LBGs at specific redshifts by sampling with different broadband filter combinations. The classic Lyman break technique has been effective at assembling large samples of LBGs in the range $z\sim3-5$ where the Lyman limit falls at optical wavelengths \citep[e.g.,][]{Steidel2003, Ouchi2004, Giavalisco2004, Verma2007, Iwata2007, Pentericci2010, Bielby2011, Oteo2013a, Alvarez2016, Malkan2017}, and the use of space-based observatories has extended the Lyman limit detection window as low as $z\sim1$ \citep[e.g.,][]{Burgarella2006, Ly2009, Basu-Zych2011, Haberzettl2012, Oteo2013b, Oteo2014, Hathi2013}.  Modified selection methods  exploiting the \lymana\ break that dominates the rest-frame UV at redshifts $z\gtrsim5$ have successfully isolated large samples of LBGs at redshifts up to $z\sim10$ \citep[e.g.,][]{Bouwens2006, Bouwens2010, Bouwens2015, McLure2011, Ellis2013, Finkelstein2016, Harikane2018, Harikane2022a}, and the redshift-dependent line blanketing by the \lya\ forest, in combination with the relatively flat rest-frame UV continuum) has been used to select LBGs in the range $1.4 < z < 2.7$ at which redshifts the Lyman limit is not observable from the ground \citep{Adelberger2004, Steidel2004}.

This feature of the Lyman break selection technique makes it particularly important  in terms of the legacy value of the current generation of deep, wide, optical and near-infrared imaging surveys.  Large-area and all-sky optical broadband photometric campaigns such as the Hyper-SuprimeCam Subaru Strategic Program \citep[HSC-SSP:][]{Aihara2018} and the imminent  Vera Rubin Observatory Legacy Survey of Space and Time \citep[LSST:][]{Ivezic2019} will exploit the Lyman break technique using 3--6 broadband filters across the rest-frame UV  to efficiently and inexpensively select hundreds of millions of galaxies in redshift ranges from $z\sim2-6$ across many hundreds to thousands of Mpc \citep[e.g.,][]{Ono2018, Harikane2018, Harikane2022a, Wilson2019}.

The Lyman break selection method comes with its own set of selection biases in favour of UV-bright, bluer star-forming galaxies with relatively low dust extinction, resulting in samples that miss a relevant fraction of UV-faint and/or heavily dust-obscured SFGs and passively evolving galaxies, particularly around the peak in cosmic star formation \citep[e.g.,][]{Grazian2007, Ly2011, Shapley2011,  Haberzettl2012, Oteo2014, Oteo2015}.  Nevertheless, LBGs are thought to dominate the UV luminosity density, and possibly the global star formation rate (SFR) density, at $z\sim 2-6$ \citep[e.g.,][]{Steidel1999, Giavalisco2004, Bouwens2009, Reddy2008, Reddy2009}, and they remain a key target population in recent surveys \citep[e.g.,][]{Arrabal2018, Ono2018, Toshikawa2018, Harikane2022b}.  Moreover, LBGs have been posited as critical populations that meet the demanding requirements of cosmological studies in the era of large-area and all-sky photometric surveys \citep[e.g.,][]{Wilson2019, Miyatake2022}, especially at higher redshifts where only methods based on \lya\ emission or Lyman break detection can be applied  in large numbers and over large scales \citep[][and references therein]{Finkelstein2016}. 

\lya\ has long been pursued as a potential tool to probe the properties of high redshift SFGs.  This endeavour has been motivated by the fact that \lya\ in absorption and/or emission is the dominant feature in the rest-frame UV spectrum of such galaxies, and is typically much stronger than other diagnostic ISM absorption and emission lines.  In addition, there are observational advantages that facilitate deep photometric imaging and spectroscopy in the wavelength range corresponding to \lya\ at $z\sim2-3$ \citep[][and references therein]{Shapley2011} -- a cosmologically critical epoch that spans the peak in SFR density \citep[][and references therein]{Madau2014} and during which more than half of the observable stellar mass of the Universe was assembled \citep[e.g.,][]{Ilbert2013,Muzzin2013}.  Moreover, \lya\ is the key -- and often the only -- observable feature in the spectra of \lya\ emitters at the highest redshifts \citep[$z\gtrsim6$,][and references therein]{Finkelstein2016, Ouchi2020} and, for this reason, has become critical for our understanding of galaxy populations during the epoch of reionisation, and their contribution to the ionising flux budget of the universe \citep[e.g.,][]{Dijkstra2014,Stark2017,Mason2018, Steidel2018}.

Due to the resonant character of the \lya\ transition, \lya\ photons are dispersed in real and frequency space whenever they encounter neutral hydrogen \citep[see][for a comprehensive description]{Dijkstra2017}.  The increased scattering and absorption experienced by \lya\ photons under the influence of these radiative transfer processes adversely affect the visibility of \lya\ emission, and complicate its spectroscopic interpretation.  However, as a direct result of these same processes, the \lya\ signal from the central few kpc of high redshift galaxies encodes information about the structure, kinematics, and ionisation properties of each galaxy and the interstellar, circumgalactic, and intergalactic media through which it propagates \citep[e.g., ][]{Shapley2003, Verhamme2006, Verhamme2008, Dijkstra2010, Steidel2010, Law2012c, Hayes2015, Trainor2015, Trainor2019, Gronke2016a, Byrohl2020, Chen2020}.

Relationships between \lya \ equivalent width (EW) and the spectral and physical properties of early SFGs have been extensively studied in populations of $z=2-4$ LBGs \citep[See for example ][]{Shapley2003, Reddy2006, Erb2006a, Law2007a, Kornei2010, Pentericci2010, Stark2010, Berry2012, Jones2012, Law2012a, Law2012c, Erb2016, Hathi2016, Trainor2016, Du2018, Marchi2019}, and especially recently in samples of the related \lymana\ emitters (LAEs) at similar redshifts \citep[e.g.,][]{Trainor2015, Trainor2019, Oyarzun2017, Guaita2017, Cullen2020, Feltre2020, Santos2020, Matthee2021}. In a systematic study at low redshift ($z\sim0.1$), the Lyman Alpha Reference Sample (LARS) collaboration investigated all the quantities thought to be involved in the \lya\ transport process \citep{Ostlin2014, Hayes2014, Pardy2014, Guaita2015, Rivera-Thorsen2015, Duval2016, Herenz2016, Runnholm2020}.  In both redshift ranges, larger \lya\ emission transmission (or \lya\ EW) was found to be associated with galaxies with bluer UV colours, lower metallicities, lower stellar masses, lower rest-frame UV luminosities, lower star formation rates, harder ionising field strengths, and more compact morphologies.  In addition, observed \lya\ emission/absorption strength has been shown to be sensitive to the galactic environment.  Not only does \lya\ visibility in the early universe reflect the well-established galaxy formation paradigm within which more luminous (massive), older \lya-absorbing LBGs occupy regions of greater mass overdensity and cluster more strongly than their lower mass, less luminous and younger LAE counterparts, \citep[e.g.,][]{Ouchi2004, Ouchi2010, Ouchi2018, Adelberger2005a, Jose2013, Bielby2016, Guaita2017}, it is also modulated by the galactic environment on small and large scales \citep[e.g.,][]{Cooke2010, Cooke2013, Diaz2014, Muldrew2015, Toshikawa2016, Lemaux2018, Shi2019, Guaita2020}.

Such relationships suggest the tantalising prospect of using \lya\ as a multi-purpose tool to elucidate the physical, environmental, and large-scale clustering properties of primordial galaxies.  To properly explore these relationships however, large samples that reflect the spectral characteristics of the selected population are necessary and, in most cases, spectroscopic measurement of \lya\ is required in order to extract the physical properties of interest.  In an approach that addressed this problem, \citet[][hereafter C09]{Cooke2009} reported a method of \lya\ spectral type classification for a population of $z\sim3$ LBGs by which pure LBG samples displaying either dominant \lya\ in absorption (aLBGs) or dominant \lya\ in emission (eLBGs) could be isolated using only broadband information.  One example of the power of this approach was demonstrated on large scales by \citet[][hereafter C13]{Cooke2013} who performed an auto- and cross-correlation function analysis of pure aLBG and eLBG samples photometrically selected from $\sim$55000 $z\sim3$ LBGs.  C13 found that aLBGs preferentially reside in group and cluster environments, eLBGs reside on the outskirts of groups and in the field, the two spectral types avoid each other on small, single halo scales, and that without accounting for the anti-correlation between aLBGs and eLBGs, masses for LBG populations were underestimated.  

One motivation for this paper is to extend the method developed by C09 to other redshifts, especially to $z\sim2$, where the availability of a statistical sample of LBGs with consistent multi-band rest-frame UV broadband photometry, uniformly measured net \lya\ EWs, and kinematic classifications quantitatively determined from IFU-based spectroscopy, prompted the investigation of the relationship between \lya\ spectral type and galaxy kinematics described in Paper\,II in this series (Foran et al.\ (2023b) submitted).

More broadly, we aim to develop a method that can be applied to large samples of $z\sim2-6$ LBGs selected from current and future large-area and all-sky photometric campaigns. The current generation of deep, large-area photometric surveys such as HSC-SSP and LSST will probe cosmic volumes and transverse areas on scales that transcend the cosmic variance of even the largest legacy fields \citep[see e.g.,][]{Arcila2013}.  The challenge for spectroscopy and targeted deep multiband/multiwavelength imaging is the huge datasets that will derive from surveys conducted on such scales; in the LSST 10-year data for example, there will be 20 billion sources that will need to be processed in order to identify hundreds of millions of high-redshift galaxies.  Optimising the discovery potential of these investments requires new techniques to statistically characterise the huge datasets they will deliver, and to efficiently select from these the most promising samples for expensive follow-up observations.  Here we explore how inexpensive broadband photometric information that is sensitive to the \lya\ properties of LBGs might be used to address these challenges, and suggest a means by which the \lya-related physical and spectroscopic properties and environments of $z\sim2-3$ LBGs might be explored on the basis of broadband photometric information alone.  

In this paper we demonstrate a statistical relationship between net \lya\ EW and the optical broadband photometric properties of $z\sim2$ LBGs.  We characterise the segregation of spectroscopic \lya-absorbing, and \lya-emitting spectral types in colour-magnitude space, and define photometric criteria by which pure sub-samples of LBGs with \lya\ dominant in absorption (p-aLBGs), and \lya\ dominant in emission (p-eLBGs), can be selected using only broadband imaging data.  As a first step toward the application of our approach to large-area and all-sky surveys, we also present here a set of ugrizy photometric selection criteria by which pure samples of p-aLBGs and p-eLBGs might be isolated from datasets derived from the LSST.

This paper is structured as follows: in Section 2, we present the photometric and spectroscopic data used in the subsequent sections. Section 3 describes the segregation versus net \lya\ EW of $z\sim2$ LBGs in colour-magnitude space, and the application of this result to determine criteria for the selection of photometric \lya\ spectral type sub-samples.  We summarise the important conclusions and potential applications of this work in Section 4.  We assume a $\Lambda$CDM cosmology with $\Omega_{M}$= 0.3, $\Omega_{\Lambda}$= 0.7 and H$_{0}$= 70\,km\,s$^{-1}$\,Mpc$^{-1}$.  All magnitudes are quoted in the AB system of \citet{Oke1983}.

\section{DATA}
\label{sec:c3_data} 

Broadband optical photometry and rest-frame net \lya\ equivalent width (hereafter `net \lya\ EW') data for a sample of 557 rest-frame UV colour-selected $z\sim2$ galaxies in the redshift range 1.7 $<z<$ 2.5 were extracted from the spectroscopic catalog of \citet{Steidel2004, Reddy2008}.  Similarly, we make use of the rest-frame net \lya\ EW measurements of \citet{Shapley2003} for a sample of 775 LBGs in the redshift range 2.5 $<z<$ 3.5 drawn from the catalog of \citet{Steidel2003}.  Values for net \lya\ EW -- which incorporate information about \lya\ in both emission and absorption -- were measured uniformly across both redshift ranges in their respective source studies using the method described by \citet{Kornei2010}.  Typical uncertainties in absolute \lya\ EW are $\sim25-50$\% for galaxies with absorption profiles, and $\sim25$\% for galaxies with \lya\ dominant in emission \citep{Shapley2003}.

The parent catalogs of the $z\sim2$ and $z\sim3$ samples derive from an observational campaign that targeted 14 uncorrelated fields with a total survey area of 1900 arcmin$^{2}$, resulting in samples that are minimally affected by systematic biases due to cosmic variance or clustering.  The survey used the $U_nG\cal{R}$ photometric system \citep{Steidel2003}, and the rest-frame UV colour selection criteria of \citet{Steidel2003} ($z\sim3$ LBGs) and \citet{Steidel2004} ($z\sim2$ BX galaxies).  These criteria were designed to recover  objects with intrinsic properties, particularly UV luminosity and reddening by dust, that were similar across both redshift ranges.  Accordingly, and although the $z\sim2$ BX selection method does not probe the Lyman break, we henceforth refer to both samples as `LBGs'.  These selection criteria result in a net \lya\ EW distribution for the $\cal{R}$ $<$ 25.5 samples that is representative of the intrinsic distribution for the parent population of galaxies \citep{Reddy2008}.  The mean redshift of our extracted $z\sim2$ sample is $z$ = 2.16 $\pm 0.20$, corresponding to a mean absolute magnitude sensitivity 0.58 mag fainter in the observed $\cal{R}$-band imaging than at $z$ = 2.96, the mean redshift of the $z\sim$3 LBG sample.  

The bulk of galaxies in the $z\sim2$ LBG sample have stellar masses in the range $9 \lesssim \mathrm{log}({M}_{\star }/{M}_{\odot }) \lesssim 11$ \citep{Shapley2005, Erb2006b, Reddy2006, Reddy2009} and star formation rates inferred from rest-frame UV luminosities (uncorrected for extinction) in the range $3 \lesssim \mathrm{M}_{\odot}$\,yr$^{-1} \lesssim 60$ \citep{Steidel2004}.  Accordingly, our $z\sim2$ sample is typical of LBGs/SFGs at these redshifts \citep[][and references therein]{Alvarez2016} and lies with a range of properties \citep[see][]{Reddy2006} on the main sequence of stellar mass and star formation rate for $z\sim2$ SFGs \citep{Daddi2007}.  

The $z\sim2$ parent sample has $\cal{R}$-band apparent magnitudes in the range $22.0 < \cal{R}$$< 25.5$, corresponding to rest-frame UV luminosities (absolute magnitudes) of $-22.6 < \mathrm{M}_{UV} < -19.1$. The faint end magnitude cut of $\cal{R}$ $\leq 25.5$ was determined by signal-to-noise requirements of the spectroscopic measurements.  Given our need for accurate $(U_n-\cal{R})$ colours, and the different $U_n$-band depths for the fields targeted by \citet{Steidel2003, Steidel2004}, we applied a further (conservative) $U_n<26.5$ cut to the $z\sim2$ , $U_n$-band data to ensure that our sample included only the most reliable photometry.  \ref{sec:photo_unc} describes the derivation of indicative photometric uncertainties for our $z\sim2$ and $z\sim3$ samples.

\section{ANALYSIS AND RESULTS}
\label{sec:c3_results}

\subsection{Net \lya\ EW distribution and spectral type classification} 
\label{sec:c3_lya_type}

The profile of \lya\ in the spectrum of high-redshift LBGs manifests in absorption, emission, or a combination of both.  In the $z\sim3$ LBG sample of \citet[][hereafter S03] {Shapley2003}, for example, the distribution of net \lya\ EWs is centred near zero and varies from $\lesssim$ $-$50 \AA\ to $\gtrsim$ $+$200 \AA.  Net \lya\ EW values for our 557 $z\sim2$ LBGs span a similar range ($-$85.0 \AA\ to $+$108.7 \AA) and, like the S03 sample, are asymmetrically dispersed toward higher net \lya\ EWs around a median near zero ($-$4.42 \AA\ at $z\sim2$ and $+$0.56 \AA\ at $z\sim3$).  These similarities, however, belie a change in the shape of the distribution that is evidenced by a shift in the mean net \lya\ EW for the respective full samples from $+$10.3 \AA\ at $z\sim3$ to $-$2.2 \AA\ at $z\sim2$.

\begin{figure}
\centering
\includegraphics[width=\columnwidth]{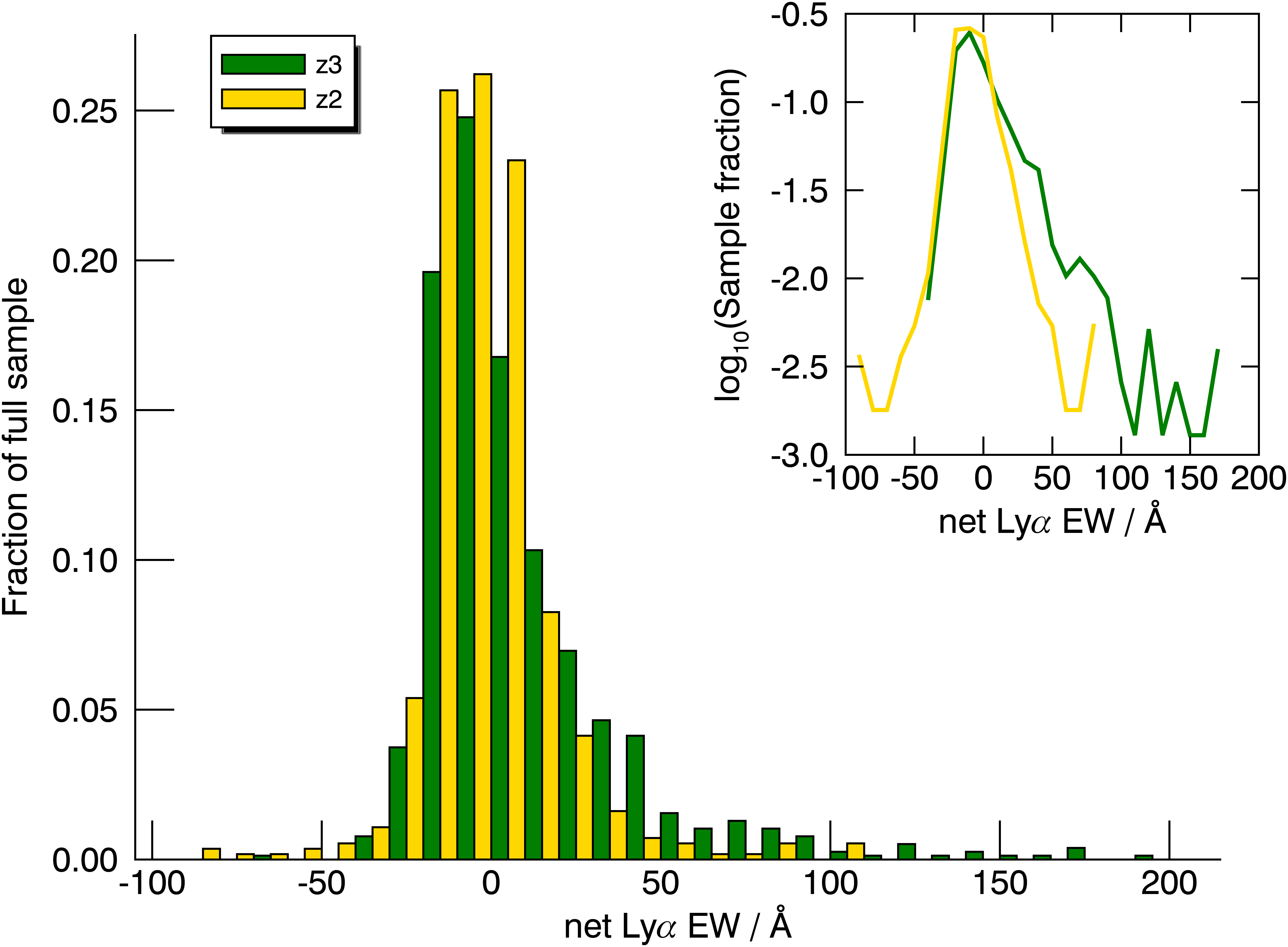}
\caption{ Normalised histograms showing the distribution of net \lya\ EWs for $z\sim3$ (green) and $z\sim2$ (gold) LBG samples.  Inset: The same distributions plotted with a logarithmic ordinate axis to accentuate the 'tails' of the \lya\ EW distributions.  Net \lya\ EWs less than zero are essentially identical between the two populations, while the $z\sim3$ sample has a significantly larger fraction of net \lya-emitters (see Table~\ref{tab:c3_table1}).}
\label{fig:c3_fig1}
\end{figure}

The changing shape of the net \lya\ EW distribution with redshift is readily apparent in Figure~\ref{fig:c3_fig1} that shows normalised histogram plots for the $z\sim2$ and $z\sim3$ samples.  Consistent with the result of \citet{Reddy2008}, we find that the two distributions are very similar at net \lya\ EWs $\lesssim$ 0 \AA, but there is a sharp drop off at $z\sim2$ toward higher values of net \lya\ EW that is largely responsible for the difference in overall mean net \lya\ EW between the two samples.

The utility of dividing a population of rest-frame UV-colour selected galaxies into sub-samples based on observed net \lya\ EW was first demonstrated by S03, and it continues to be a useful approach in the study of relationships between \lya\ and the physical and spectral properties of LBGs \citep[e.g., ][]{Du2018, Pahl2020}.  In their discovery of the broadband photometric segregation versus \lya\ EW in the S03 sample, C09 exploited the same approach to derive the method of photometric \lya\ spectral-type classification (see Section~\ref{sec:c3_seg}.

Table~\ref{tab:c3_table1} shows a comparison of the statistics for our $z\sim2$ and $z\sim3$ LBG samples divided into numerical quartiles on the basis of net \lya\ EW.  There is a shift toward more negative mean net \lya\ EW ($\Delta \sim -5$\,\AA) of the most absorbing quartile at $z\sim2$ compared to the same quartile at $z\sim3$.  The two $z\sim2$ quartiles that span the more \lya-emitting end of the distribution show a larger (and increasing) shift to lower average net \lya\ EW compared to the analogous quartiles of S03 ($\Delta -9.0$\,\AA\ and $\Delta -32.5$\,\AA\ for q3 and q4 respectively).

Motivated by these observations, and our results showing a relationship between net \lya\ EW and nebular emission-line kinematics (Foran et al.\ (2023b) submitted), we applied to our $z\sim2$ sample the same net \lya\ EW cuts used by S03 to generate numerical quartiles at $z\sim3$. We define the most absorbing fraction of galaxies with net \lya\ EW $\leq$ $-$10.0\,\AA\ as `aLBGs', and the most strongly emitting fraction with net \lya\ EW $\geq$ $+$20.0\,\AA\ as `eLBGs'.  We further divide the remaining LBGs into $\rm{G_a}$ and $\rm{G_e}$ spectral types with net \lya\ EWs $-$10.0\,\AA\ $<$ net \lya\ EW $<$ 0.0\,\AA\ and 0.0\,\AA\ $<$ net \lya\ EW $<$ $+$20.0\,\AA\ respectively.  Table~\ref{tab:c3_table1} summarises the population statistics of the $z\sim2$ sample and our \lya\ spectral types compared to the $z\sim3$ LBGs of S03.

\begin{table*}
\centering
\caption{Statistics for sub-samples of $z\sim2$ and $z\sim3$ LBGs divided on the basis of net \lya\  EW}
\label{tab:c3_table1}
\begin{threeparttable}
\begin{tabular}{ccccccccc}
\toprule
\multicolumn{9}{c}{\thead{Redshift range}} \\ 
\multicolumn{3}{c}{\thead{$z\sim3$\tnote{a}}} 
& \multicolumn{3}{c}{\thead{$z\sim2$\tnote{b}}} 
& \multicolumn{3}{c}{\thead{$z\sim2$\tnote{c}}} 
\\
\thead{$N$\tnote{d}} 
& \thead{$N$/$N_{tot}$\tnote{e}} 
& \thead{\lya\ EW\tnote{f} \\ (\AA)} 
& \thead{$N$} 
& \thead{$N$/$N_{tot}$} 
& \thead{\lya\ EW \\ (\AA)} 
& \thead{$N$} 
& \thead{$N$/$N_{tot}$} 
& \thead{\lya\ EW \\ (\AA)} 
\\
\cmidrule(lr){1-3} \cmidrule(lr){4-6} \cmidrule(lr){7-9}
775 & 1.0 & +10.3 & 557 & 1.0 & -2.2 & 557 & 1.0 & -2.2 \\
\cmidrule(lr){1-3} \cmidrule(lr){4-6} \cmidrule(lr){7-9}
194 & 0.25 & -16.7 & 140 & 0.25 & -21.5 & 188 & 0.34 & -18.9  \\
193 & 0.25 & -4.8 & 139 & 0.25 & -8.6 & 146 & 0.26 & -5.3  \\
194 & 0.25 & +8.3 & 139 & 0.25 & +0.7 & 176 & 0.32 & +6.9   \\
194 & 0.25 & +54.3 & 139 & 0.25 & +20.8 & 47 & 0.08 & +40.6 \\
\bottomrule
\end{tabular}
\begin{tablenotes}
\item [a] $z\sim3$ LBGs from  \citet{Shapley2003}
\item [b]  Our $z\sim2$ sample divided into numerical quartiles
\item [c]  Our $z\sim2$ sample divided into aLBG, $\rm{G_a}$, $\rm{G_e}$ and eLBG \lya\ spectral types as per the definitions given in Section~\ref{sec:c3_lya_type}.
\item [d]  Number of galaxies
\item [e]  Fraction of full sample ($N_{tot}$) in sub-sample ($N$).
\item [f]  Mean net \lya\ EW for each (sub)sample
\end{tablenotes}
\end{threeparttable}
\end{table*}

Given that we have defined our spectral types using the same net \lya\ EW cuts as S03, it is not surprising that the central ($\rm{G_a}$ and $\rm{G_e}$) spectral types have mean net \lya\ EWs similar ($\Delta \sim -1$\,\AA) to the equivalent quartiles (q2 and q3) in the $z\sim3$ sample.  It is noteworthy, however, that despite the overall shift in mean net \lya\ EW of $\sim -12.5$\,\AA\ between the $z\sim3$ and $z\sim2$ samples, the mean net \lya\ EW of the $z\sim2$ aLBGs is similarly only $\sim -2$\,\AA\ more negative than the equivalent S03 quartile, indicative of a compression of the LBG population toward the \lya-absorbing end of the distribution.  This behaviour is likely due to the fact that the measured net \lya\ EW becomes insensitive to the total absorption once the \lya\ absorption feature is saturated.  That is, beyond that point, any further decrease in measured EW would reflect only the contribution of the damping wings, and depend weakly on increasing \HI\ column density (see Section~\ref{sec:c3_seg} for manifestation of this effect in the broadband imaging data).  

Conversely, the mean net \lya\ EW of the eLBG spectral type sub-sample at $z\sim2$ is \ $\sim -14$\,\AA\ more negative than the analogous (most strongly \lya-emitting) S03 quartile, and the relative fraction of eLBGs at $z\sim2$ is 0.08 compared to 0.25 at $z\sim3$ -- a clear reflection of the lower relative abundance of net \lya\ emitting LBGs in the universe and/or within the LBG selection function at $z\sim2$ compared to $z\sim3$.

\subsection{Segregation of $z\sim$ 2 LBGs in colour-magnitude space} 
\label{sec:c3_seg}

\begin{figure*}
\centering
\scalebox{0.43}[0.43]{\includegraphics{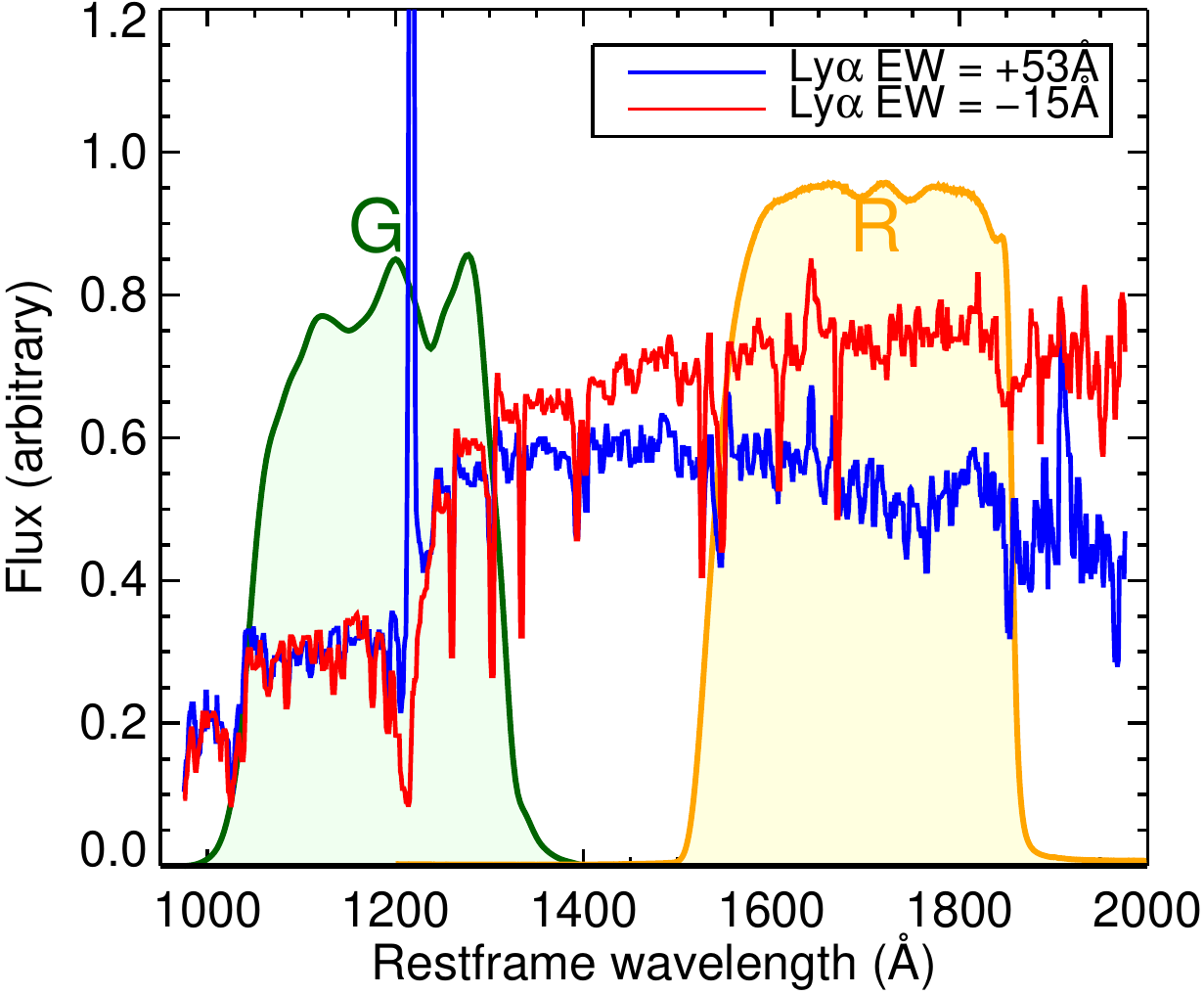}}
\scalebox{0.43}[0.43]{\includegraphics{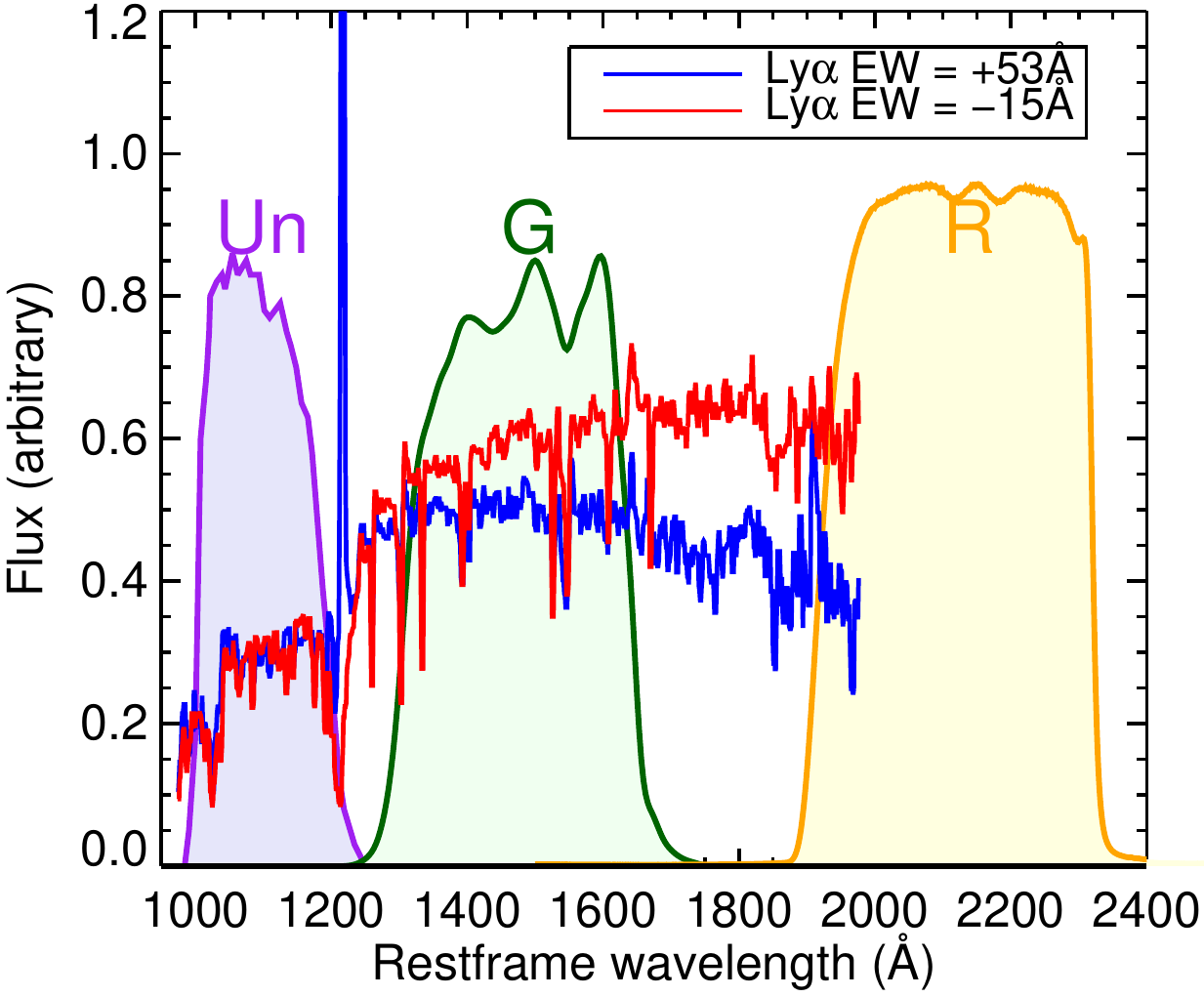}}
\caption{Illustration of the origin of the colour separation of Lyman break galaxy (LBG) \lya\ spectral types. Left: Plotted are the $G$ and $\cal{R}$ filter transmission curves \citep{Steidel2003} in green and orange, respectively, shifted to the $z\sim3$ rest-frame. Overlaid are the (smoothed) quartile 1 (red, representative of aLBGs) and quartile 4 (blue, representative of eLBGs) composite spectra of \citet{Shapley2003}. The composite spectra consist of $\sim$200 $z\sim3$ LBG spectra with similar \lya\ EW, with the mean values indicated in the legend. The spectra are shown normalised over the ${G}$ filter to help illustrate the ($G - \cal{R}$) colour difference in the two spectral types for a given ${G}$ magnitude. The origin of the \lya\ spectral type photometric segregation on the $(G-\cal{R})$ vs.\ $\cal{R}$ CMD results from their colour differences based on the UV continuum slope relationship with spectral type and a small (and inverse) contribution from the \lya\ emission/absorption feature and the magnitude differences in spectral type, in that $z\sim3$ aLBGs are brighter on average than eLBGs.  Right: Similar to the left plot, but for LBGs at z $\sim$ 2.  The composite spectra are shown normalised over the $U_n$ filter (violet, see text).  Note: the composite spectra and the normalisation are shown for illustrative purposes and extend to 2000\AA, rest-frame. However, the UV continuum slopes of quartiles 1 (red) and 4 (blue) maintain a significant difference in $\cal{R}$ that is sufficient to separate aLBG and eLBG spectral types in ($U_n - \cal{R}$) colour and $\cal{R}$ magnitude on the CMD. Depending on the redshift of z $\sim$ 2 LBGs, the \lya\ feature may fall in or out of the $U_n$ filter (see Section~\ref{sec:c3_lya}).}
\label{fig:c3_fig2}
\end{figure*}

C09 discovered that over the redshift path $z \sim 3.0 \pm 0.3$, the relationship between rest-frame UV continuum slope and net \lya\ EW leads to a photometric dispersion of LBGs, and an ability to separate LBG spectral types on a broadband colour-magnitude plane based on their net \lya\ EW. At $z\sim3$, aLBGs are (on average) brighter in $\cal{R}$ magnitude and redder than eLBGs.  They separate in $(G-\cal{R})$ colour as a result of the redder UV continuum slopes of aLBGs, combined with an additional small red enhancement as a result of the \lya\ absorption in the $G$-band, as compared to eLBGs with bluer UV continuum slopes, combined with an additional blue enhancement from the \lya\ emission in the $G$-band. Together, these behaviours enable a statistical segregation of the two populations on a $(G-\cal{R})$ vs.\ $\cal{R}$ colour-magnitude diagram (CMD), with subsets containing pure samples of each spectral type.  Figure~\ref{fig:c3_fig2} illustrates the origin of the $z\sim3$ broadband imaging segregation of \lya-absorbing and \lya-emitting LBGs that enables the determination of photometric \lya\ spectral types.  

To test whether a similar relationship between broadband photometry and net \lya\ EW might exist at $z\sim2$, we use $(U_n-\cal{R})$ colours and $\cal{R}$-band magnitudes to construct a CMD for our sample of 557 $z \sim2$ LBGs.  We use $(U_n-\cal{R})$ rather than $(U_n-G)$, to sample the rest-frame UV continuum farther redward of the \lya\ feature so as to increase the segregation between the redder-sloped aLBGs and the bluer-sloped eLBGs, and to avoid any possibility of contamination of our redward filter by \lya\ emission.  The separation in wavelengths probed by the $U_n$ and $G$ filters at $z\sim2$ is smaller than the separation of the $G$ and $\cal{R}$ filters at $z\sim3$ (see Figure~\ref{fig:c3_fig2}). 

\begin{figure*}
\centering
\scalebox{0.58}[0.58]{\includegraphics{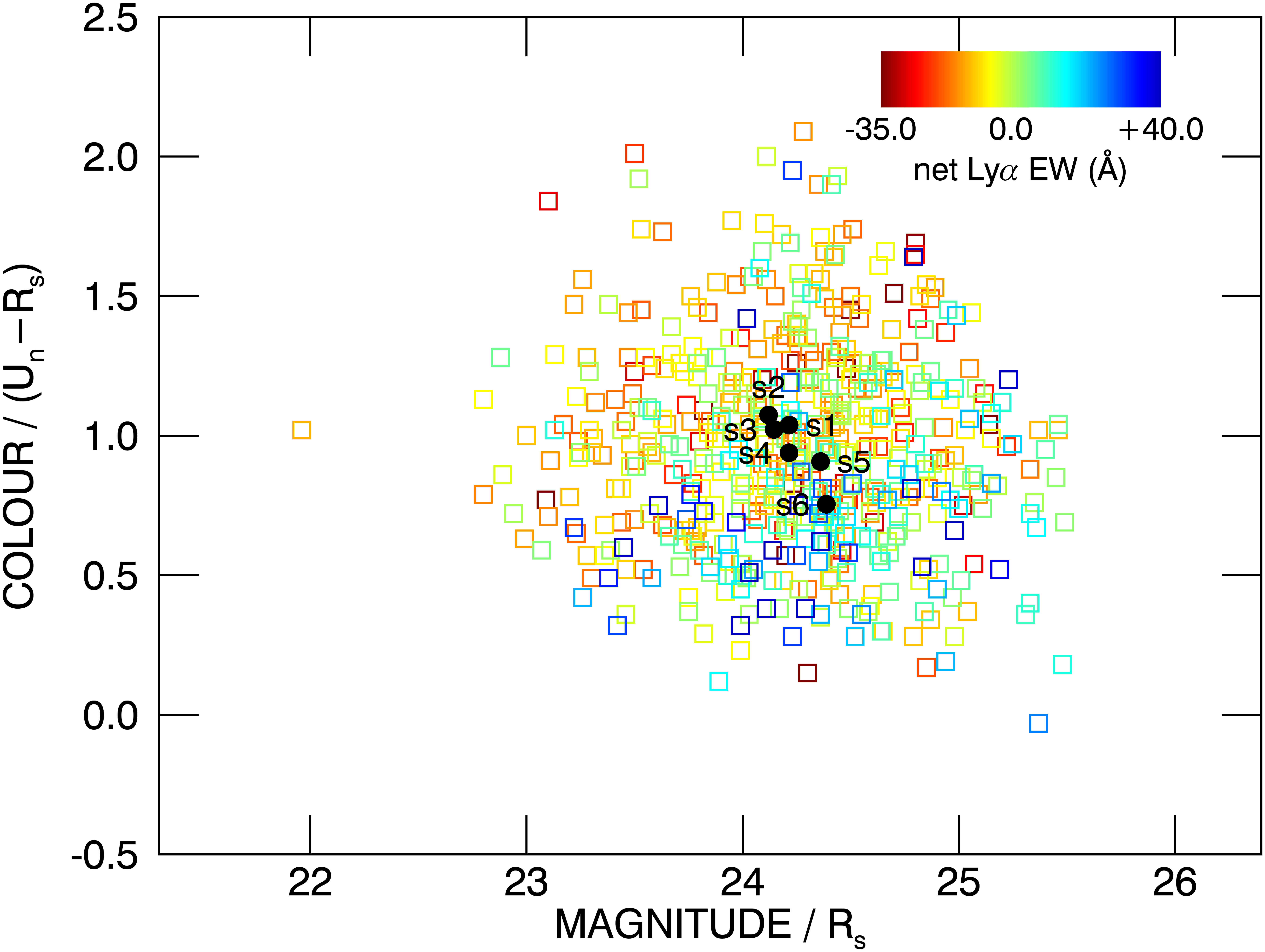}}
\scalebox{0.58}[0.58]{\includegraphics{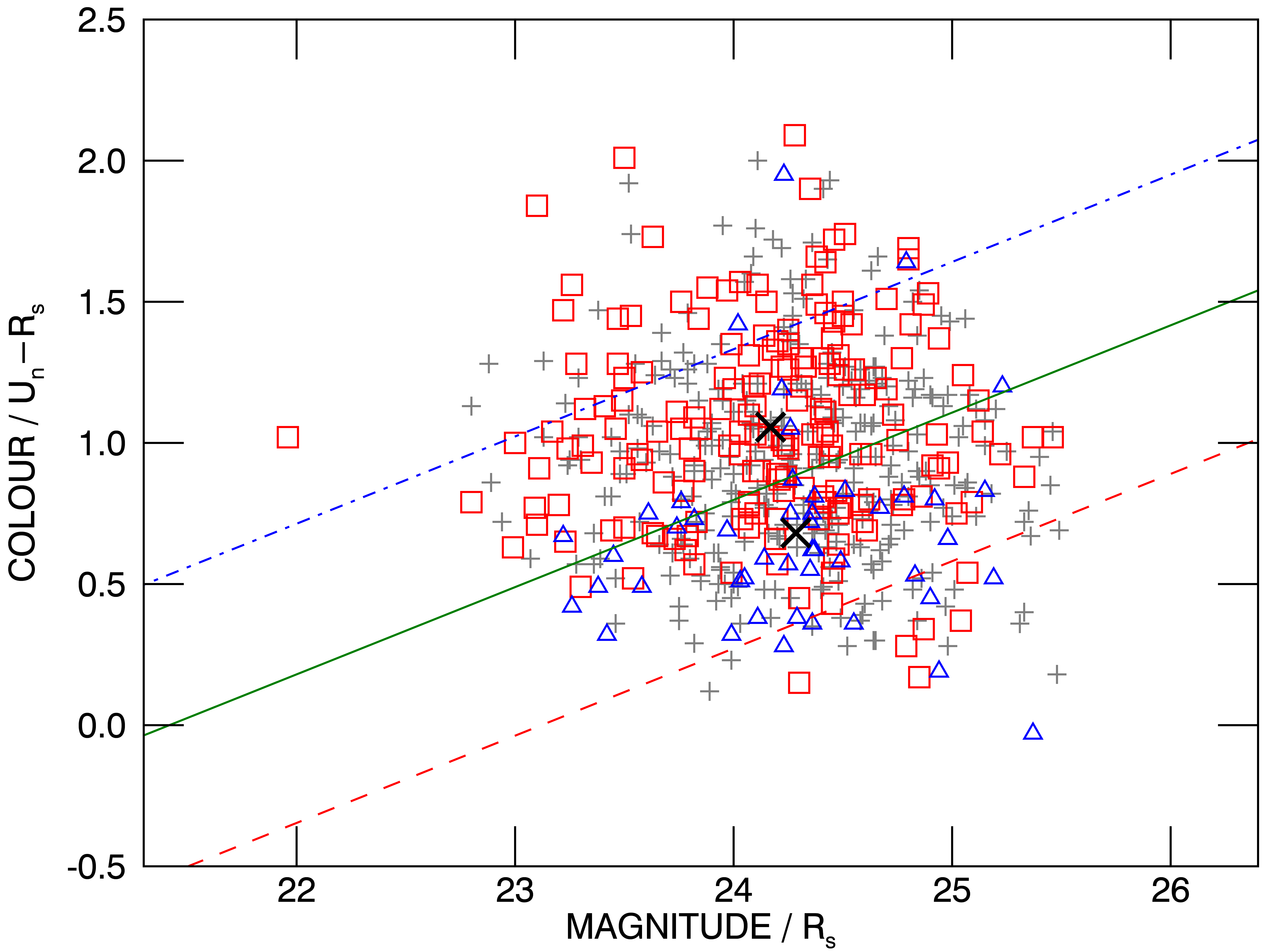}}
\caption{ Rest-frame UV colour $(U_n-\cal{R})$--magnitude ($\cal{R}$) diagrams (CMDs) for Lyman break galaxies (LBGs) in the redshift range $1.7<z<2.5$, and with magnitude cuts of $\cal{R}$ $<$ 25.5 and $U_n$ $<$ 26.5. Left: $z\sim2$ LBGs dispersed in colour-magnitude space with symbols colour-coded on a red-blue gradient according to their measured net \lya\ EW.  The colour table maps the range $-$35.0\,\AA\ $<$ net \lya\ EW $<$ $+$40.0\,\AA, which encompasses $\gtrsim$ 95\% of the sample. Points labelled s1 to s6 indicate the colour and magnitude distribution means of the LBG sample numerical sextiles divided on the basis of net \lya\ EW. Right: Grey plus (+) marks denote the 557 galaxies in the $z\sim2$ spectroscopic sample. Galaxies with net \lya\ EW $\leq -10.0$\,\AA\ (aLBGs) are overlaid with red squares, and those with net \lya\ EW $\geq +20.0$\,\AA\ (eLBGs) are overlaid with blue triangles. The mean value for each distribution is marked with a black cross (X), with aLBG mean indicated by the upper cross and eLBG mean by the lower. The dotted-dashed blue and dashed red lines indicate a 1.5$\sigma$ dispersion in colour from the primary cut (green line) that divides the aLBG and eLBG distributions, respectively (see text).}
\label{fig:c3_fig3}
\end{figure*}

The left panel of Figure~\ref{fig:c3_fig3} shows the spectroscopic $z\sim2$ LBG sample dispersed in colour $(U_n-\cal{R})$ and magnitude $\cal{R}$, with symbols colour-coded on a red-to-blue gradient according to their measured net \lya\ EW. For visualisation purposes the colour table is scaled to map the range $-$35.0\,\AA\ $<$ net \lya\ EW $<$ $+$40.0\,\AA\, which encompasses $\gtrsim$ 95\% of the galaxies in our sample. Plotting the data in this way (i.e., downplaying the colour effect of the few extreme net \lya\ EW cases), the bulk of the LBG sample manifests on the CMD as a visible colour gradient from red to blue moving diagonally from roughly the top left to bottom right.  The galaxies in our sample with the most negative net \lya\ EW (dark red symbols)  do not lie at the extreme end of the colour gradient direction as might be expected for a simple monotonic relationship. While they are certainly well within the `absorbing' half of the CMD, they lie toward the centre of the distribution, and approximately along a line orthogonal to the underlying trend. This apparently anomalous behaviour of the most absorbing galaxies in our sample notwithstanding, the overall trend is confirmed by the points labelled s1 to s6 on the colour gradient plot, that indicate the positions of the magnitude and colour distribution means for the numerical sextiles (of $\sim$93 galaxies each) grouped according to their net \lya\ EW (see Table \ref{tab:table2} for a summary of the sextile statistics).  The more positive net \lya\ EW LBGs (weaker absorption and more emission) show an overall trend toward fainter $\cal{R}$-band magnitudes and bluer ($U_n-\cal{R}$) colours.  Indeed, only the most absorbing sextile (s1 in Figure~\ref{fig:c3_fig3}) does not follow this monotonic trend. That being said, the colours and magnitudes of the $z\sim2$ sextiles converge with increasing \lya\ absorption strength (i.e., from s6 to s1), unlike at $z\sim3$, where the mean colours (magnitudes) continue to redden (brighten) monotonically (cf.\ C09).

\begin{table} \begin{center} 
\caption{Statistics for the dispersion of $z\sim2$ LBGs in colour ($U_n-\cal{R}$) -- magnitude ($\cal{R}$) space divided into numerical sextiles based on net \lya\ EW} 
\label{tab:table2} 
\begin{threeparttable}
\begin{tabular}{cccc} 
\toprule 
\thead{Sextiles\tnote{a}} &
\thead{Mean \\ $\cal{R}$ mag.} & \thead{Mean \\ ($U_n-\cal{R}$) colour} &
\thead{Mean \\ \lya\ EW (\AA)}  
\\ 
\midrule 
\thead{s1} & 24.22 & 1.04 & -25.08\\ 
\thead{s2} & 24.12 & 1.07 & -12.85\\ 
\thead{s3} & 24.15 & 1.02 & -7.21\\ 
\thead{s4} & 24.21 & 0.94 & -0.86\\ 
\thead{s5} & 24.36 & 0.91 & 5.66\\
\thead{s6} & 24.39 & 0.75 & 27.44\\ 
\bottomrule 
\end{tabular}
\begin{tablenotes} 
\item {\bf [a]  Full $z\sim2$ spectroscopic LBG sample (557 galaxies) divided into sextiles of $\sim$93 galaxies each.}  
\end{tablenotes}
\end{threeparttable} \end{center} \end{table}

We can speculate that the `off trend' positions of the strongest \lya\ absorbers on the CMD is a manifestation of the environmental effect proposed by C13 over their intrinsic net \lya\ EWs. In this scenario, the most negative net \lya\ EWs observed, with otherwise typical aLBG colour and magnitude, may be a result of their environment near the cores of groups and proto-clusters and the presence of larger column densities of intra-group/cluster neutral gas at or near the systemic velocity of the LBGs along the line of sight \citep[e.g.,][]{Muldrew2015, Toshikawa2016, Lemaux2018}. More prosaically, it is also plausible that this behaviour, and the grouping of the three most absorbing sextiles (s1--s3) on the CMD, is a reflection of the compression of the $z\sim2$ sample towards the more negative end of the net \lya\ EW distribution (as described in Section~\ref{sec:c3_lya_type}), combined with the inherently greater uncertainties (25--50\%) associated with the net \lya\ EW measurements for the most absorbing systems (S03) and the inherent photometric scatter (see~\ref{sec:photo_unc}).

The association between net \lya\ EW and the photometric properties of $z\sim2$ LBGs suggested by the trend shown in the left panel of Figure~\ref{fig:c3_fig3}, prompts a statistical examination using the \lya\ spectral type classification scheme described in Section \ref{sec:c3_lya_type} and the method demonstrated by C09.  In the right panel of Figure~\ref{fig:c3_fig3} we plot on the CMD the LBG spectral types as described in Section~\ref{sec:c3_lya_type}, i.e., aLBGs with net \lya\ EW $\le$ -10 \AA\ and eLBGs with net \lya\ EW $\ge$ $+$20 \AA, and show that they segregate into two cohesive, albeit overlapping, distributions.

We define a primary cut (solid green line) that passes through the midpoint between the mean colour and magnitude values of the aLBG and eLBG distributions (black crosses), and has slope that maximises the difference in mean net \lya\ EW and spectral-type purity between the sub-samples that lie above and below the broken blue and red lines respectively.  These dashed (red) and dotted-dashed (blue) lines indicate an offset of 1.5$\sigma$ in colour dispersion from the primary cut for the aLBG and eLBG distributions respectively (see Section~\ref{sec:c3_z2_purity}).  Statistics for the segregation of the aLBG and eLBG spectral types shown in the right panel of Figure~\ref{fig:c3_fig3} are summarised in Table~\ref{tab:c3_table3} together with (for comparison) the segregation statistics for the $z\sim3$ sample of C09.

\begin{table}
\centering
\caption{Statistics for the photometric segregation of \lya-absorbing and \lya-emitting spectral types in $z\sim$ 2 and $z\sim$ 3 LBGs}
\label{tab:c3_table3}
\begin{threeparttable}
\begin{tabular}{lcccc}
\toprule
& \multicolumn{4}{c}{\thead{Redshift}} \\
& \multicolumn{2}{c}{\thead{$z\sim2$}} 
& \multicolumn{2}{c}{\thead{$z\sim3$}}  
\\
\midrule
\thead{Colour}
& \multicolumn{2}{c}{$(U_n-\cal{R})$} 
& \multicolumn{2}{c}{$(G-\cal{R})$} 
\\
\thead{Magnitude}
& \multicolumn{2}{c}{$\cal{R}$} 
& \multicolumn{2}{c}{$\cal{R}$} 
\\
\thead{Total LBGs} 
& \multicolumn{2}{c}{557} 
& \multicolumn{2}{c}{775}  
\\
\midrule
& \thead{aLBGs} & \thead{eLBGs} 
 & \thead{aLBGs} & \thead{eLBGs} \\
      \midrule
\thead{Number} 
& 188 & 47 
& 144 & 150
\\ 
\thead{Magnitude mean} 
& 24.17 & 24.28 
& 24.43  & 24.94
\\
\thead{Colour mean} 
& 1.06 & 0.68 
& 0.78 & 0.40
\\
\thead{Colour $\sigma$} 
& 0.35 & 0.36 
& 0.24 & 0.31
\\
\thead{Slope} 
& \multicolumn{2}{c}{0.31} 
& \multicolumn{2}{c}{0.40}
\\
\thead{Intercept} 
& \multicolumn{2}{c}{-6.62} 
& \multicolumn{2}{c}{-9.38} 
\\
\bottomrule
    \end{tabular}
\end{threeparttable}
\end{table}

Although the slope and intercept for the $z\sim2$ segregation quoted in Table~\ref{tab:c3_table3} are the values that give the maximum difference in mean net \lya\ EW between the photometrically-selected sub-samples, the maximum is shallow, asymmetric, and relatively insensitive to the choice of slope.  For example, in the optimal case where $c_{\sigma} = 1.25$ (see Section~\ref{sec:c3_z2_purity}), the maximum difference in mean net \lya\ EW is 17.1\,\AA\ at a slope of 0.31.  We note, however, that the difference in mean net \lya\ EW is greater than 16.0\,\AA\ for all slopes between 0.17 and 0.40.  Thus we might quote an uncertainty (or `range of confidence') of slope = $0.31^{+0.09}_{-0.13}$.  within which any choice of slope would result in photometrically-selected sub-samples with a difference in mean net \lya\ EW that is within $\sim$5\% of the maximum.  Constraining the primary cut to pass through the mid-point of the aLBG and eLBG distribution means similarly gives intercept values in the range $-6.62^{+3.43}_{-2.30}$.

\subsection{Photometric \lya\ spectral type selection and sub-sample purity}
\label{sec:c3_purity}

\subsubsection{$z\sim2$ LBGs}
\label{sec:c3_z2_purity}

Following the method used by C09 at $z\sim3$, we use the parameters of the segregated aLBG and eLBG distributions shown in Figure~\ref{fig:c3_fig3} to isolate sub-samples with \lya\ dominant in absorption (`photometric' aLBGs, or p-aLBGs) and with \lya\ dominant in emission (`photometric' eLBGs, or p-eLBGs) from the parent $z\sim2$ LBG sample.  Invoking the primary cut slope and intercept values from Table~\ref{tab:c3_table3}, and the supplied broadband photometry, the following relationships can be used to extract sub-samples of the desired photometric \lya\ spectral type.

For p-aLBGs, 
\begin{equation}
\label{eq:c3_albgs} 
\mbox{($U_n - \cal{R}$)} ~ \ge ~
\mbox{0.3091} \cdot \cal{R} - \mbox{6.6208} ~ \mbox{+ c$_\sigma$} \cdot
\mbox{$\sigma_e$} 
\end{equation}

\noindent and for p-eLBGs, 
\begin{equation}
\label{eq:c3_elbgs} 
\mbox{($U_n - \cal{R}$)} ~ \le ~ 
\mbox{0.3091} \cdot \cal{R} - \mbox{6.6208} ~ 
\mbox{- c$_\sigma$} \cdot \mbox{$\sigma_a$} 
\end{equation}

\noindent where $c_{\sigma}$ is the coefficient of colour standard deviation by which boundaries used to isolate the photometric spectral type sub-samples are offset from the primary cut on the CMD, and ${\sigma}_{a}$ (0.3509) and ${\sigma}_{e}$ (0.3558) are the 1$\sigma$ standard deviations of the $(U_n-\cal{R})$ colour distributions for the aLBG and eLBG subsets, respectively. 

We use ${\sigma}_a$ and ${\sigma}_e$ to estimate the density of aLBGs and eLBGs on the CMD.  This approach implies that any $\sigma$ should extend around a distribution mean in some circular (or similar) contour.  The primary cut we make between the aLBG and eLBG distribution means (and its use as the basis for estimating photometric spectral type purity) is a line for which our assumptions only formally apply at the point of closest approach (tangent to a circular contour) of our lines to the respective distribution means.  Thus, the multiples of $c_{\sigma}$ (1.5 in Figure~\ref{fig:c3_fig3}) applied to ${\sigma}_a$ and ${\sigma}_e$, plus the fraction of $\sigma$ by which the primary cut is removed from the respective distribution mean positions ($\sim 0.5 {\sigma}_a$ and $\sim 0.5 {\sigma}_e$), represent a minimum coefficient of $\sigma$ that can be used to estimate the extent and purity of different \lya\ spectral types on the CMD.  For example, cuts on the CMD for which $c_{\sigma} = 1.5$ along the same slope as the primary cut approximate (for Gaussian distributions) criteria for selecting \lya\ spectral type sub-samples $\gtrsim$ 2$\sigma$ from the mean value of the opposite distribution.  

In theory, the above criteria can be made stricter (or relaxed) by varying the value of $c_{\sigma}$, thereby trading sub-sample size for sub-sample purity according to the properties of the parent sample, and the requirements of the intended application.  In practice, the range of $c_{\sigma}$ values that can be meaningfully employed is limited by the degree to which the aLBG and eLBG distributions deviate from Gaussian behaviour, and by small-number statistics at higher values of $c_{\sigma}$ -- especially for eLBGs which are $\sim$4 times less abundant than aLBGs in our $z\sim2$ sample.  Table~\ref{tab:c3_table4} summarises the statistics for p-aLBG and p-eLBG \lya\ spectral type sub-samples selected from the parent $z\sim2$ LBGs using the selection criteria given in Equations~\eqref{eq:c3_albgs} and \eqref{eq:c3_elbgs}, respectively, and a range of $c_{\sigma}$ values.

We estimate the purity of each photometric spectral type sub-sample by calculating the degree to which they exclude galaxies with opposite spectral type as determined by their measured net \lya\ EW and our classification scheme described in Section~\ref{sec:c3_lya_type}. That is, for example, for each p-aLBG sub-sample selected using a different value of $c_{\sigma}$, we calculate the contamination fraction of spectroscopic eLBGs and $\rm{eLBG} + \rm{G_e}$ spectral types. The purity of the p-aLBG sub-sample thus determined is quoted as a percentage with respect to eLBGs and with respect to $\rm{eLBG}$ + $\rm{G_e}$ spectral types (parenthesised) in Table~\ref{tab:c3_table4}.  The mean net \lya\ EW of the p-aLBG and p-eLBG sub-samples (also listed in Table~\ref{tab:c3_table4}) is a further measure of the quality of the broadband photometric segregation, and the average properties of the respective sub-samples.

Across a wide range of $c_{\sigma}$ values, we select high-purity p-aLBG sub-samples, particularly with respect to contamination by spectroscopic eLBGs.  Indeed, even using the primary cut between the aLBG and eLBG spectral types (i.e., $c_{\sigma} = 0.0$), results in a large sub-sample of 311 p-aLBGs ($\gtrsim$ 55\% of total LBGs) that is $\gtrsim$ 97\% free of eLBGs and $\gtrsim$ 70\% pure with respect to galaxies with any detectable net \lya\ emission.  The practical upper limit of $c_{\sigma}$ for the selection of p-aLBGs appears to be restricted only by the diminishing return of smaller sub-sample sizes. As a result, large broadband photometric samples can greatly benefit from stricter cuts. The optimal coefficient for the dataset here of $\sigma_e$ ($c_{\sigma} \approx 1.0 -1.25$) selects $\sim 100 - 140$ photometric aLBGs that are $\gtrsim$ 97\% and $\gtrsim$ 79\% pure with respect to spectroscopic eLBGs, and $\rm{eLBG}$ + $\rm{G_e}$ spectral types respectively, and for which the mean net \lya\ EW is $\sim$ $-$8\,\AA. 

With $c_{\sigma} \approx 1.0 - 1.25$ we select a sample of $\sim 50 - 70$ LBGs with \lya\ dominant in emission (p-eLBGs) that are $\gtrsim 85 \%$ and $\gtrsim 65 \%$ pure with respect to spectroscopic aLBGs and $\rm{aLBG}$ + $\rm{G_a}$ spectral types respectively, with a mean net \lya\ EW of $\sim$ +8\,\AA.  These purities represent a significant enhancement over the native (full sample) abundances of eLBGs ($\sim$8\%) and the sum of eLBG and $\rm{G_e}$ spectral types ($\sim$40\%).  

The segregation versus net \lya\ EW of photometrically selected $z\sim2$ p-aLBG and p-eLBG sub-samples with $c_{\sigma} = 1.0$ is plotted in the top panel of Figure~\ref{fig:c3_fig4} compared to the distribution versus net \lya\ EW of the parent $z\sim2$ LBG sample.  In the optimal case, we select sub-samples with a desired \lya\ spectral type at $z\sim2$ that are for p-aLBGs, comparable to, and for p-eLBGs $\sim 10\%$ less pure than, the optimised $z\sim3$ result of C09 (see Section~\ref{sec:c3_z3_purity}).  The lower optimised purity of p-eLBGs at $z\sim2$ is attributable to the intrinsic overlap of the aLBG and eLBG distributions, and the relatively lower fraction of \lya-emitting LBGs selected at this redshift.  That is, the ratio of aLBGs to eLBGs has increased from around 1:1 at $z\sim$3 to more than 4:1 at $z\sim$2 when comparing samples to the same absolute magnitude.  This is not specific to our sample, On the contrary, a reduced fraction of \lya-emitting galaxies with decreasing redshift is expected from the findings of \citet{Stark2010, Stark2011, Mallery2012, Cassata2015} who consistently report an evolutionary decrease in the fraction of LBGs with \lya\ in emission from $z\sim6$ to $z\sim2$ at fixed luminosity. 

\begin{table}
\begin{center}
\caption{Statistics for photometric sub-samples with \lya\ dominant in absorption (p-aLBGs) and \lya\ dominant in emission (p-eLBGs) selected from the parent $z\sim2$ LBG sample using Equations~\eqref{eq:c3_albgs}~\&~\eqref{eq:c3_elbgs} and different values of $c_{\sigma}$.}
\label{tab:c3_table4}
\begin{threeparttable}
\begin{tabular}{ccccccc}
\toprule
& \multicolumn{6}{c}{\thead{Photometric Spectral Type}} \\ 
 & \multicolumn{3}{c}{\thead{p-aLBGs}} 
& \multicolumn{3}{c}{\thead{p-eLBGs}}
\\
\thead{$c_{\sigma}$\tnote{a}}
& \thead{$N$\tnote{b}} 
& \thead{Purity\tnote{c} \\ (\%)} 
& \thead{\lya\ EW\tnote{d} \\ (\AA)} 
& \thead{$N$} 
& \thead{Purity \\ (\%)} 
& \thead{\lya\ EW \\ (\AA)} 
\\
\midrule
0.0 & 311 & 97.1(70.7) & -6.25 & 246 & 76.0(53.6) & 3.03 \\
0.5 & 223 & 98.2(73.1) & -7.59 & 156 & 80.1(62.2) & 5.06 \\
0.75 & 178 & 97.8(74.7) & -8.00 & 104 & 84.6(66.3) & 7.03 \\
1.0 & 138 & 97.8(79.0) & -8.23 & 69 & 85.5(65.2) & 7.73  \\
1.25 & 101 & 97.0(79.2) & -7.42 & 50 & 84.0(64.0) & 9.59 \\
1.5 & 72 & 95.8(77.8) & -5.80 & 36 & 83.3(61.1) & 2.94  \\
2.0 & 38 & 97.4(76.3) & -7.19 & 11 & 54.5(45.4) & -4.35 \\
\bottomrule
   \end{tabular}
\begin{tablenotes}
\item [a] Coefficient of colour standard deviations ($\sigma_a$~\&~$\sigma_e$) by which boundaries used to isolate the photometric spectral type sub-samples are offset from the CMD primary cut
\item [b] Number of galaxies in the photometric sub-samples
\item [c] For p-aLBGs: Percent purity with respect to eLBG and (eLBG + $\rm{G_e}$) spectral types.  For p-eLBGs: Percent purity with respect to aLBG and (aLBG + $\rm{G_a}$) spectral types.  
\item [d] Mean net \lya\ EW for each sub-sample
\end{tablenotes}
\end{threeparttable}
 \end{center}
\end{table}

\subsubsection{$z\sim3$ LBGs}
\label{sec:c3_z3_purity}

For the purposes of reference and direct comparison, we present here the \lya\ spectral type photometric selection results for $z\sim3$ LBGs analysed and presented in the same format as the $z\sim2$ result above.

Parameters for the photometric segregation of $z\sim3$ LBG \lya-absorbing and \lya-emitting spectral types in $(G-\cal{R})$ vs $\cal{R}$ colour-magnitude space as determined by C09 are listed in Table~\ref{tab:c3_table3}, and we re-produce in Equations~\eqref{eq:c3_z3_albgs}~\&~\eqref{eq:c3_z3_elbgs} criteria for the photometric selection of p-aLBG and p-eLBG spectral type sub-samples at $z\sim3$.

For p-aLBGs, 
\begin{equation}
\label{eq:c3_z3_albgs} 
\mbox{($G - \cal{R}$)} ~ \ge ~
\mbox{0.4047} \cdot \cal{R} - \mbox{9.3760} ~ \mbox{+ c$_\sigma$} \cdot
\mbox{$\sigma_e$} 
\end{equation}

\noindent and for p-eLBGs, 
\begin{equation}
\label{eq:c3_z3_elbgs} 
\mbox{($G - \cal{R}$)} ~ \le ~ \mbox{0.4047} \cdot 
\cal{R} - \mbox{9.3760} ~ \mbox{- c$_\sigma$} \cdot \mbox{$\sigma_a$} 
\end{equation}

\noindent where for the parent sample of $z\sim3$ LBGs, $\sigma_a = 0.2392$ and $\sigma_e = 0.3095$.

\begin{figure} \centering
\includegraphics[width=\columnwidth]{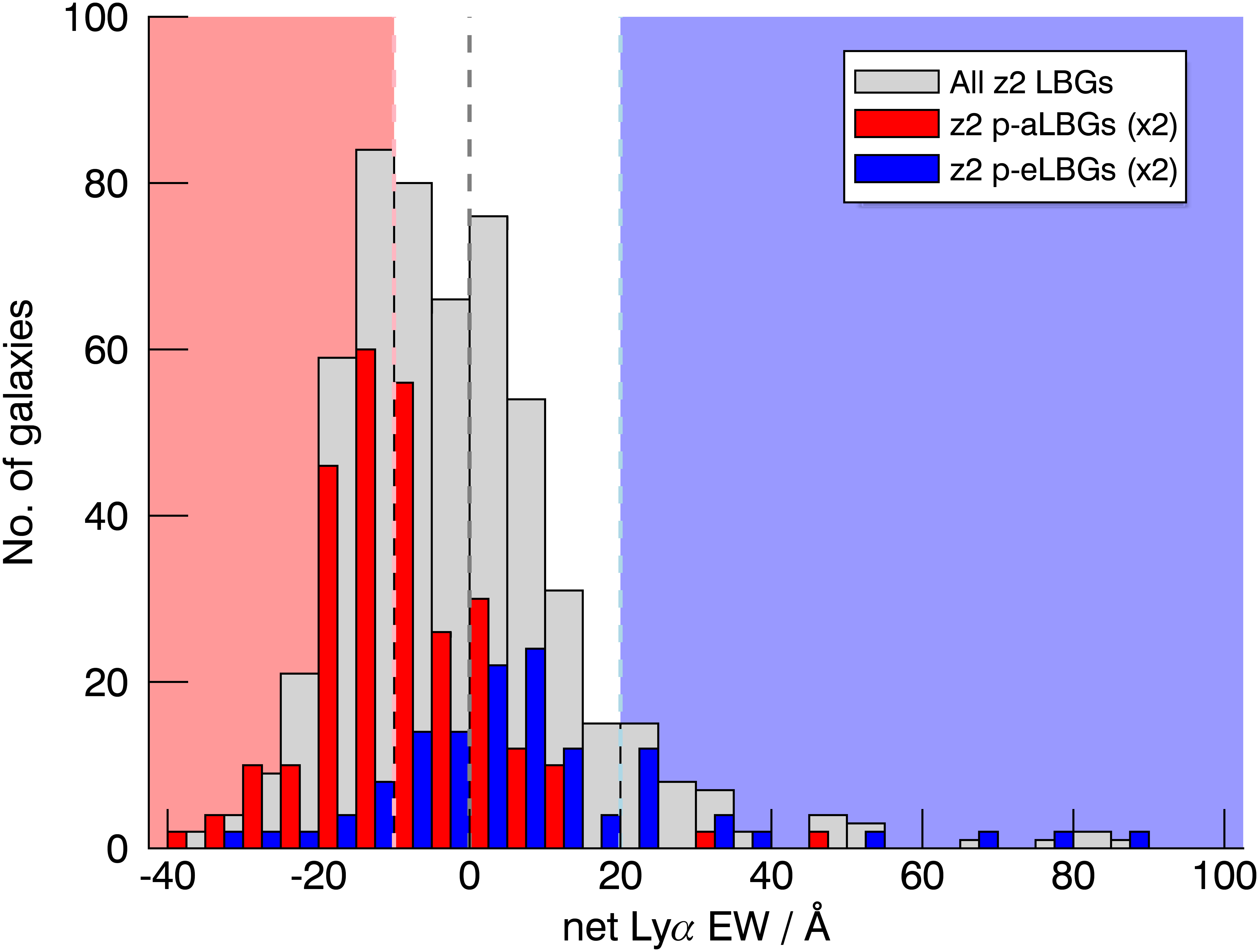}
\includegraphics[width=\columnwidth]{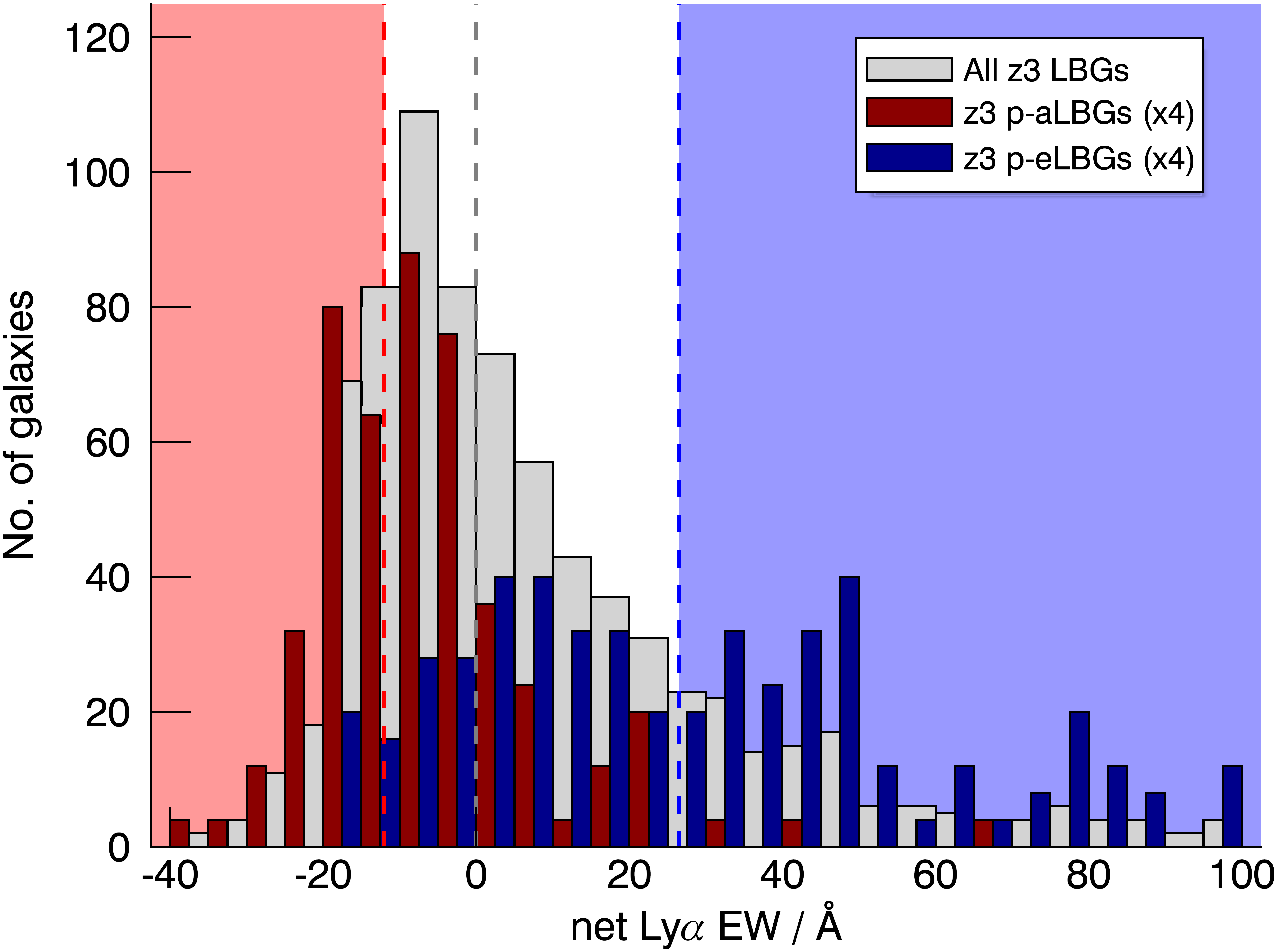}
\caption{Histograms versus net \lya\ EW of $z\sim2$ and $z\sim3$ `photometric' aLBG (p-aLBG) and `photometric' eLBG (p-eLBG) spectral type sub-samples overlaid on the distribution versus net \lya\ EW of their respective parent samples shown in grey.  Vertical dashed lines indicate the net \lya\ EW thresholds used here to divide the spectroscopic sample into aLBG, $\rm{G_a}$, $\rm{G_e}$ and eLBG \lya\ spectral types.  Red and blue shaded regions indicate aLBGs and eLBGs respectively.  Top: $z\sim2$ p-aLBGs and p-eLBGs selected from the parent sample of 557 $z\sim2$ LBGs using the selection criteria given in Equations~\eqref{eq:c3_albgs}~\&~\eqref{eq:c3_elbgs} with $c_{\sigma}=1.0$.  Histograms of p-aLBGs and p-eLBGs are multiplied by 2 for clarity.  Bottom: $z\sim3$ p-aLBGs and p-eLBGs selected from the parent sample of 775 $z\sim3$ LBGs using the selection criteria given in Equations~\eqref{eq:c3_z3_albgs}~\&~\eqref{eq:c3_z3_elbgs} with $c_{\sigma}=1.5$.} Histograms of p-aLBGs and p-eLBGs are multiplied by 4 for clarity.
\label{fig:c3_fig4} 
\end{figure}

Table~\ref{tab:c3_table5} summarises the statistics for p-aLBG and p-eLBG \lya\ spectral type sub-samples selected from a parent sample of 775 $z\sim3$ LBGs using the above relationships and a range of $c_{\sigma}$ values.  The bottom panel of Figure~\ref{fig:c3_fig4} shows the segregation versus net \lya\ EW of $z\sim3$ p-aLBG and p-eLBG sub-samples selected with $c_{\sigma} = 1.5$ compared to the distribution versus net \lya\ EW of the parent $z\sim3$ LBG sample.  

The optimal coefficient of $\sigma_e$ ($c_{\sigma} \approx 1.5$) selects $\sim120$ photometric aLBGs that are $\gtrsim$ 96\% and $\gtrsim$ 76\% pure with respect to spectroscopic eLBGs, and $\rm{eLBG}$ + $\rm{G_e}$ spectral types respectively, and for which the mean net \lya\ EW is $\sim$ $-$5\,\AA. 

With any coefficient of $\sigma_a \gtrsim 1.0$, we select large samples of photometric eLBGs that are $\sim 94-98$\% pure with respect to spectroscopic aLBGs.  Over the range $c_{\sigma} = 1.0-2.5$, the purity of the p-eLBG sample with respect to all net \lya-absorbers ($\rm{aLBG}$ + $\rm{G_a}$ spectral types) increases monotonically from $\sim$74\% to $\sim$93\%, with a commensurate increase in mean net \lya\ EW from $\sim+27$ to $\sim+$51~\AA.

\begin{table}
\begin{center}
\caption{Statistics for photometric sub-samples with \lya\ dominant in absorption (p-aLBGs) and \lya\ dominant in emission (p-eLBGs) selected from the parent sample of 775 $z\sim3$ LBGs using the spectral type criteria of C09\tnote{a} and different values of $c_{\sigma}$.}
\label{tab:c3_table5}
\begin{threeparttable}
\begin{tabular}{ccccccc}
\toprule
& \multicolumn{6}{c}{\thead{Photometric Spectral Type}} \\ 
& \multicolumn{3}{c}{\thead{p-aLBGs}} 
& \multicolumn{3}{c}{\thead{p-eLBGs}}
\\
\thead{$c_{\sigma}$\tnote{b}}
& \thead{$N$\tnote{c}} 
& \thead{Purity\tnote{d} \\ (\%)} 
& \thead{\lya\ EW\tnote{e} \\ (\AA)} 
& \thead{$N$} 
& \thead{Purity \\ (\%)} 
& \thead{\lya\ EW \\ (\AA)} 
\\
\midrule
0.0 & 419 & 92.6 (64.4) & 0.0 & 356 & 91.0 (69.1) & $+$22.4 \\
1.0 & 200 & 94.5 (70.5) & $-$2.9 & 192 & 94.3 (74.5) & $+$27.3 \\
1.5 & 119 & 96.6 (76.5) & $-$5.3 & 131 & 95.4 (82.4) & $+$33.7 \\
2.0 & 56 & 98.2 (73.2) & $-$6.4 & 91 & 94.5 (82.4) & $+$37.7  \\
2.5 & 20 & 95.0 (65.0) & $-$7.3 & 55 & 98.2 (92.7) & $+$51.4  \\ 
\bottomrule
   \end{tabular}
\begin{tablenotes}
\item [a] For the purposes of determining photometric segregation criteria, C09 defined aLBGs and eLBGs as having net \lya\ EW $\leq -12.0$ and $\geq +26.5$\,\AA\ respectively.
\item [b] Coefficient of colour standard deviations ($\sigma_a$~\&~$\sigma_e$) by which boundaries used to isolate the photometric spectral type sub-samples are offset from the CMD primary cut
\item [c] Number of galaxies in the photometric sub-samples
\item [d] For p-aLBGs: Percent purity with respect to eLBG and (eLBG + $\rm{G_e}$) spectral types.  For p-eLBGs: Percent purity with respect to aLBG and (aLBG + $\rm{G_a}$) spectral types.  
\item [e] Mean net \lya\ EW for each sub-sample
\end{tablenotes}
\end{threeparttable}
 \end{center}
\end{table}

\subsection{The contribution of \lya}
\label{sec:c3_lya}

The segregation of $z\sim3$ aLBGs and eLBGs on the CMD is enhanced by the contribution of the \lya\ feature itself when it falls within the bandpass of the $G$ filter (see Section~\ref{sec:c3_seg} and Figure~\ref{fig:c3_fig2}).  The contribution of \lya\ to the observed luminosities was estimated by C13 to be $\sim$-0.1 mags for aLBGs, $\sim$+0.1 mags for eLBGs , and negligible for LBGs with net \lya\ EW near zero.  A similar effect might be anticipated at $z\sim2$ when \lya\ falls within the bandpass of the relevant ($U_n$) filter.

Unlike the $z\sim3$ case where the \lya\ spectral feature lies within the $G$-band filter across the full redshift range of the sample ($2.5<z<3.5$), about half ($\sim$52\%) of the $z\sim2$ LBG sample is in the redshift range $2.17 \leq z \leq 2.50$, where the \lya\ feature lies outside the half-power bandpass limits of the $U_n$ filter.  \citet{Reddy2008} showed that the ratio of strong emitters to absorbers for LBGs at redshifts $2.17 \leq z \leq 2.48$ is approximately the same as for those selected by the same set of colour criteria at $z<2.17$ (see the respective population statistics in Table~\ref{tab:c3_table6}).  Thus, there is no underlying selection bias of aLBGs versus eLBGs that could affect the segregation properties in the different redshift ranges. This result does not, however, preclude the possibility that the segregation statistics across the full $z$-range of the sample may be variably affected by the contribution of \lya\ to the measured $U_n$-band photometry. For this reason -- and because the measured segregation at $z\sim2$ is less well resolved than at $z\sim3$ -- we look to quantify the effect of \lya\ on the observed broadband segregation for galaxies in the $z\sim2$ sample in different redshift ranges and with different \lya\ spectral type.  

To this end, we divide the $z\sim2$ LBG sample into two subsets: one containing only galaxies in the range $1.7<z<2.17$ where \lya\ falls within the bandpass of the $U_n$ filter (3250--3850\,\AA), and another comprising galaxies in the $2.17<z<2.5$ range for which \lya\ lies beyond the red half-power bandpass limit of the same filter (see Figure~\ref{fig:c3_fig2}).  We then optimise the primary cut slope in each redshift bin in the same manner as for the sample as a whole (see Section~\ref{sec:c3_seg}), and compare the segregation statistics for the two subsets with each other, and with those for the full $z$-range sample (see Table~\ref{tab:c3_table6}).  

We find a significantly stronger segregation in the $2.17<z<2.5$ sample, most apparent in the greater average colour segregation between aLBGs and eLBGs in this redshift range (0.52) compared to that in the lower redshift bin (0.24).  This effect is likely due to the larger contribution of the \lya\ forest to the$U_n$-band of the $2.17<z<2.5$ sample, leading to redder $(U_n-\mathcal{R})$ colours on average.  There is also a larger colour dispersion of eLBGs ($\sigma = 0.41$) in the lower redshift range that blurs the photometric segregation.  

This difference in the degree of segregation between aLBGs and eLBGs translates into the purity of p-aLBG and p-eLBG sub-samples that can be selected from the two redshift ranges.  Using the same methodology as was applied to the full $z\sim2$ and $z\sim3$ samples in Section~\ref{sec:c3_purity}, we determined the purity of p-aLBG and p-eLBG sub-samples selected from each redshift bin using the segregation parameters listed in Table~\ref{tab:c3_table6}, and a range of $c_{\sigma}$ values see Table~\ref{tab:c3_table7}).  The optimised purity of the p-aLBGs and p-eLBGs selected from the $2.17<z<2.5$ sample ($\sim98$\% and $\sim94$\% respectively) is significantly better than can be achieved from the lower redshift bin ($\sim94$\% for p-aLBGs and $\sim84$\% for p-eLBGs).  In fact, with $c_{\sigma}$ values of 1.0 to 1.5, the p-aLBG and p-eLBG sub-samples in the $2.17<z<2.5$ range have purities that are essentially indistinguishable from those achievable in the $z\sim3$ sample, and comparably high $\Delta_{Ly\alpha\ EW}$ between them, indicating that any direct contribution of the \lya\ feature to the segregation properties of aLBGs and eLBGs is dominated by other redshift-dependent spectrophotometric effects such as the one described above.

\begin{table*}
\centering
\caption{Statistics for the segregation of $z\sim2$ LBGs in color ($U_n-\cal{R}$) -- magnitude ($\cal{R}$) space over different redshift ranges}
\label{tab:c3_table6}
\begin{threeparttable}
\begin{tabular}{ccccccc}
\toprule
& \multicolumn{6}{c}{\thead{Redshift range}} \\ 
& \multicolumn{2}{c}{\thead{$1.7<z<2.5$}} 
& \multicolumn{2}{c}{\thead{$1.7<z<2.17$}} 
& \multicolumn{2}{c}{\thead{$2.17<z<2.5$}} \\
\cmidrule(rl){1-1}  \cmidrule(rl){2-7} 
\thead{Total LBGs} 
& \multicolumn{2}{c}{557} 
& \multicolumn{2}{c}{266} 
& \multicolumn{2}{c}{291} 
\\
\thead{$z$ mean} 
& \multicolumn{2}{c}{2.158} 
& \multicolumn{2}{c}{1.983} 
& \multicolumn{2}{c}{2.318} 
\\
\thead{$\mathcal{R}$ mean} 
& \multicolumn{2}{c}{24.24} 
& \multicolumn{2}{c}{24.18} 
& \multicolumn{2}{c}{24.29} 
\\
\thead{$(U_n-\mathcal{R})$ mean} 
& \multicolumn{2}{c}{0.96} 
& \multicolumn{2}{c}{0.83} 
& \multicolumn{2}{c}{1.07} 
\\
\thead{Net \lya\ EW mean \\ (\AA)} 
& \multicolumn{2}{c}{$-$2.15} 
& \multicolumn{2}{c}{$-$3.85} 
& \multicolumn{2}{c}{$-$0.60} 
\\
\cmidrule(lr){2-7} 
& \thead{aLBGs} & \thead{eLBGs} 
& \thead{aLBGs} & \thead{eLBGs} 
& \thead{aLBGs} & \thead{eLBGs} 
\\
\cmidrule(rl){1-1} \cmidrule(lr){2-3} \cmidrule(lr){4-5} \cmidrule(lr){6-7}
\thead{Number} 
& 188 & 47 
& 96 & 20 
&92 & 27
\\ 
\thead{$\mathcal{R}$ mean} 
& 24.17 & 24.28 
& 24.13 & 24.21 
& 24.21 & 24.34
\\ 
\thead{$(U_n-\mathcal{R})$ mean} 
& 1.06 & 0.68 
& 0.91 & 0.67 
& 1.21 & 0.69
\\ 
\thead{Colour $\sigma$} 
& 0.35 & 0.36
& 0.31 & 0.41 
& 0.32 & 0.32
\\ 
\thead{Net \lya\ EW mean \\ (\AA)} 
& $-$18.87 & $+$40.57
& $-$19.84 & $+$38.13 
& $-$17.86 & $+$42.38
\\
\thead{Slope} 
& \multicolumn{2}{c}{ 0.31} 
& \multicolumn{2}{c}{0.11} 
& \multicolumn{2}{c}{0.16} 
\\
\thead{Intercept} 
& \multicolumn{2}{c}{ -6.62 } 
& \multicolumn{2}{c}{-1.75} 
& \multicolumn{2}{c}{-2.89} 
\\
\bottomrule
    \end{tabular}
\end{threeparttable}
\end{table*}

\begin{table*}
\centering
\caption{Statistics for p-aLBG \& p-eLBG sub-samples photometrically selected from the parent $z\sim2$ LBG sample using segregation parameters optimised in different redshift ranges}
\label{tab:c3_table7}
\begin{threeparttable}
\begin{tabular}{cccccccc}
\toprule
\multicolumn{8}{c}{\thead{Redshift range : $1.70 < z < 2.17$}} \\ 
\cmidrule(rl){2-4} \cmidrule(lr){5-7} 
 & \multicolumn{3}{c}{\thead{p-aLBGs}} & \multicolumn{3}{c}{\thead{p-eLBGs}} & \\
\thead{$c_{\sigma}$\tnote{a}} & \thead{$N$\tnote{b}} & \thead{Purity\tnote{c} \\ (\%)} & \thead{\lya\ EW\tnote{d} \\ (\AA)} & \thead{$N$} & \thead{Purity \\ (\%)} & \thead{\lya\ EW \\ (\AA)} & \thead{$\Delta_{Ly\alpha\ EW}$\tnote{e}} \\
\cmidrule(rl){1-1} \cmidrule(rl){2-4} \cmidrule(lr){5-7} \cmidrule(rl){8-8} 
0.00 & 142 & 96.5 (71.8) & -7.5 & 126 & 71.4 (46.8) & 0.1 & 7.6 \\
0.25 & 113 & 96.5 (71.7) & -6.8 & 97 & 79.4 (54.6) & 1.5 &  8.3 \\
0.50 & 82 & 95.1 (69.5) & -6.7 & 75 & 82.7 (56.0) & 2.8 & 9.4 \\
0.75 & 50 & 94.0 (62.0) & -6.4 & 57 & 84.2 (57.9) &  3.8 & 10.2 \\
1.00 & 32 & 93.8 (75.0) & -9.7 & 38 & 81.6 (44.7) &  0.4 & 10.1 \\
1.25 & 21 & 90.5 (71.4) & -4.6 & 23 & 78.3 (52.2) & 1.0 & 5.6 \\
1.50 & 13 & 84.6 (61.5) & 0.0 & 16 & 68.8 (43.8) &  -1.1 & 1.1 \\
2.00 &  5 & 100.0 (80.0) & -14.5 & 6 & 66.7 (66.7) & -3.5 & 11.1 \\
\midrule
\midrule
\multicolumn{8}{c}{\thead{Redshift range : $2.17 < z < 2.50$}} \\ 
\cmidrule(rl){2-4} \cmidrule(lr){5-7} 
 & \multicolumn{3}{c}{\thead{p-aLBGs}} & \multicolumn{3}{c}{\thead{p-eLBGs}} & \\
\thead{$c_{\sigma}$} & \thead{$N$} & \thead{Purity \\ (\%)} & \thead{\lya\ EW \\ (\AA)} & \thead{$N$} & \thead{Purity \\ (\%)} & \thead{\lya\ EW \\ (\AA)} & \thead{$\Delta_{Ly\alpha\ EW}$} \\
\cmidrule(rl){1-1} \cmidrule(rl){2-4} \cmidrule(lr){5-7} \cmidrule(rl){8-8} 
0.00 & 174 & 98.9 (70.7) & -6.0 & 115 & 80.9 (63.5) & 7.7 & 13.7 \\
0.25 & 157 & 98.7 (72.6) & -6.5 & 98 & 84.7 (67.3) &  9.7 & 16.2 \\
0.50 & 131 & 99.2 (77.1) & -8.2 & 76 & 84.2 (73.7) & 11.1 & 19.3 \\ 
0.75 & 107 & 99.1 (75.7) & -7.9 & 52 & 90.4 (80.8) & 13.8 & 21.8 \\
1.00 & 84 & 98.8 (78.6) & -7.8 & 31 & 93.5 (87.1) & 19.9 & 27.7 \\
1.25 & 65 & 98.5 (81.5) & -8.6 & 19 & 94.7 (94.7) & 19.4 & 28.0 \\
1.50 & 49 & 98.0 (79.6) & -8.2 & 12 & 100.0 (100.0) & 25.3 & 33.5 \\
2.00 & 25 & 96.0 (76.0) & -6.0 & 2 & 100.0 (100.0) & 10.5 & 16.4 \\
\bottomrule
   \end{tabular}
\begin{tablenotes}
\item [a] Coefficient of colour standard deviations by which boundaries used to isolate the photometric spectral type sub-samples are offset from the primary cut in each redshift range
\item [b] Number of galaxies in the photometric sub-samples
\item [c] For p-aLBGs: Percent purity with respect to eLBG and (eLBG + $\rm{G_e}$) spectral types.  For p-eLBGs: Percent purity with respect to aLBG and (aLBG + $\rm{G_a}$) spectral types.  
\item [d] Mean net \lya\ EW for each sub-sample
\item [e]  Difference in mean net \lya\ EW between the p-aLBG and p-eLBG sub-samples. In the case of Full$_{bins}$, ${\Delta}_{Ly\alpha\ EW}$ is the weighted average of $\Delta_{Ly\alpha\ EW}$ values for each magnitude bin.
\end{tablenotes}
\end{threeparttable}
\end{table*}

\subsection{LSST photometric selection criteria for $z\sim3$ LBG \lya\ spectral types}
\label{sec:lsst}

A key objective of this work is to develop a method that can be applied to large samples of $z\sim2-6$ LBGs identified from current and future large-area and all-sky photometric campaigns. As a first step toward this goal, we adapt the spectrophotometric method of C13 to model photometric selection criteria by which populations of $z\sim3$ LBGs with \lya\ dominant in absorption and \lya\ dominant in emission might be selected from the broadband \ugri\ photometric data of the Vera Ruben Observatory Legacy Survey of Space and Time (VRO/LSST). 

\begin{figure} 
\centering
\includegraphics[width=\columnwidth]{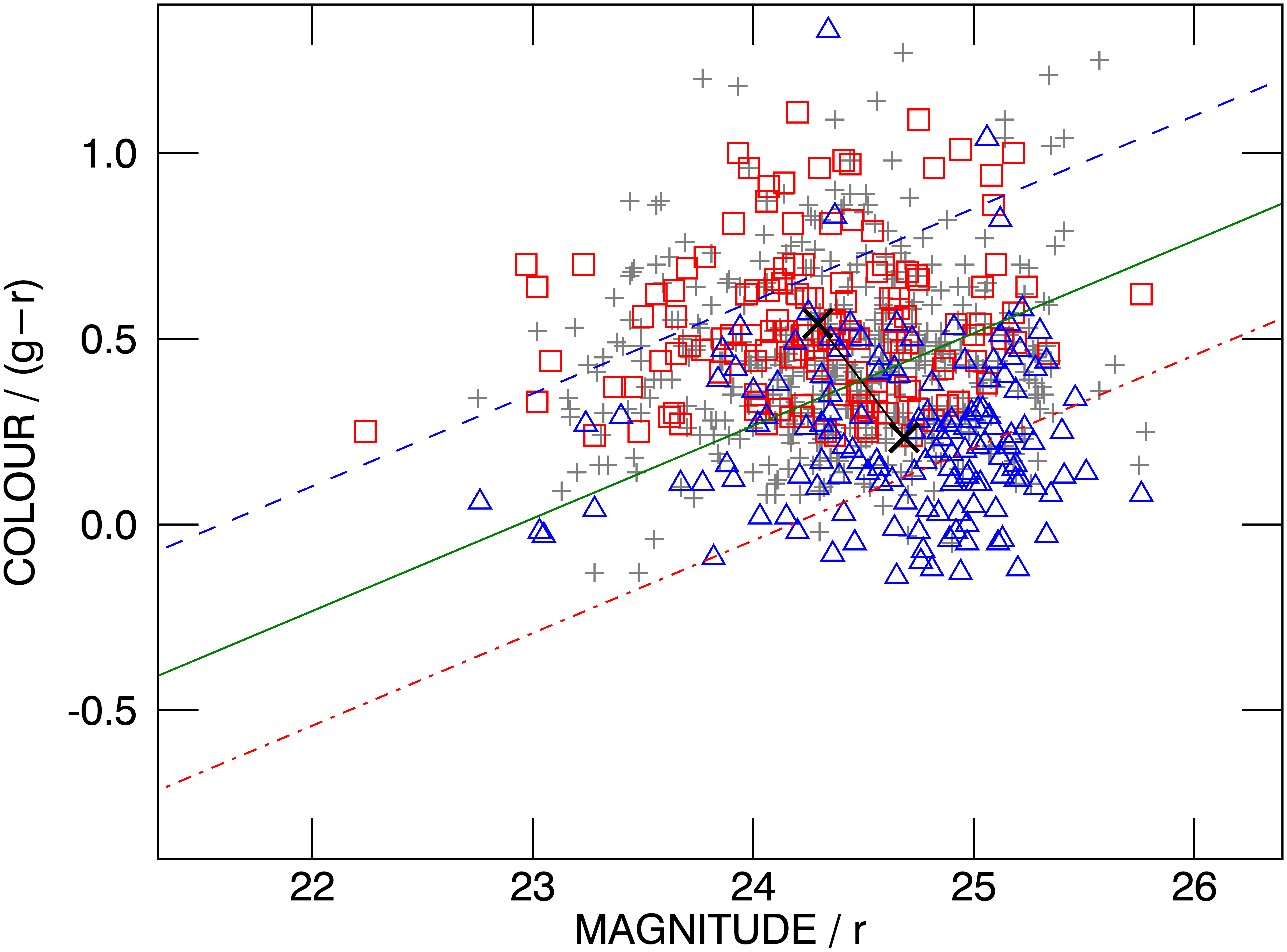}
\caption{Rest-frame UV $(g-r)$ vs.\ $r$ colour-magnitude diagram (CMD) for \citet{Steidel2003} $z\sim3$ LBGs, constructed with VRO/LSST \ugri\ photometry derived from the net \lya\ EW quartile composite spectra of \citet{Shapley2003}.  The CMD shows the segregation with net \lya\ EW of galaxies with \lya\ dominant in absorption (aLBGs, red squares) and \lya\ dominant in emission (eLBGs, blue triangles) Grey plus (+) symbols denote galaxies with intermediate values of net \lya\ EW.  Black crosses mark the mean positions of the aLBG and eLBG distributions.  The dashed red and dotted-dashed blue lines indicate a $1.5\sigma$ dispersion in colour from the primary cut (green line) for the aLBG and eLBG distributions, respectively.}
\label{fig:c3_fig8} 
\end{figure}

In order to model the segregation statistics for different \lya\ spectral types in the LSST \ugri\ photometric system, it is first necessary to calculate \ugri\ magnitudes for each galaxy in our $z\sim3$ sample.  Our photometric segregation method relies on the fact that LBGs with different net \lya\ EW have different spectral properties -- in particular rest-frame UV continuum slope -- that give rise to different rest-frame UV colours depending on their \lya\ absorbing/emitting properties (see Figure~\ref{fig:c3_fig2}).  Thus, in order to convert from \UGR\ to \ugri\ magnitudes via spectrophotometry, we must be able to assign to each galaxy in our sample, an appropriate  spectrum corresponding to its \lya\ spectral type. For this purpose, we make use of the four composite spectra of \citet{Shapley2003}, derived from the $z\sim3$ \UGR\ LBGs described in Sections~\ref{sec:c3_data}~\&~\ref{sec:c3_lya_type} divided into quartiles on the basis of net \lya\ EW.  C13 showed that spectrophotometry of these composite spectra accurately reproduces the magnitude and colour means and dispersions of each of the four net \lya\ EW quartile samples, as well as the full distribution of $z\sim3$ LBGs on the $(G-\cal{R})$/$G$ CMD when combined.  Thus, although the colours and magnitudes of individual galaxies vary within each quartile, the composite spectra can be used to compute net \lya\ EW means and dispersions on the CMD for our $z\sim3$ LBG sample when viewed through the VRO /LSST \ugri\ filters\footnote{LSST filter bandpasses and throughputs (31 May 2021 updates) downloaded from: \url{https://github.com/lsst/throughputs/tree/main/baseline}}. Figure~\ref{fig:c3_fig8} shows our $z\sim3$ \UGR\ LBG sample dispersed in VRO/LSST $(g-r)$ vs.\ $r$ colour--magnitude space with photometry thus derived from spectrophotometry of the net \lya\ EW quartile composite spectra.

\begin{table}
\begin{center}
\caption{Statistics for the segregation of LSST $z\sim3$ LBG \lya\ spectral types in $(g-r)$/$r$ colour--magnitude space}
\label{tab:lsst_table1}
\begin{threeparttable}
\begin{tabular}{ccc}
\toprule
& \multicolumn{2}{c}{\thead{\lya\ Spectral Type}}\tnote{a} \\ 
& \thead{aLBGs} & \thead{eLBGs} \\
\midrule
\thead{Mag.\ mean} 
& 24.29 & 24.68 \\
\thead{Mag.\ $1\sigma$}
& 0.54 & 0.54 \\
\thead{Col.\ mean} 
& 0.54 & 0.23 \\
\thead{Col.\ $1\sigma$}
& 0.21 & 0.22 \\
\thead{Slope} 
& \multicolumn{2}{c}{0.25} \\
\thead{Intercept} 
& \multicolumn{2}{c}{-5.72} \\
\bottomrule
    \end{tabular}
\begin{tablenotes}
\item [a] Consistent with C09, we define aLBGs and eLBGs as having net \lya\ EW $\leq -12.0$ and $\geq +26.5$\,\AA\ respectively, for the purposes of constructing the CMD and determining photometric segregation criteria.
\end{tablenotes}
\end{threeparttable}
  \end{center}
\end{table}

Following the method described in Section~\ref{sec:c3_purity}, we use the parameters of the segregated aLBG and eLBG distributions shown in Figure~\ref{fig:c3_fig8} and tabulated in Table~\ref{tab:lsst_table1} to determine photometric selection criteria by which pure populations of $z\sim3$ LBGs with \lya\ dominant in absorption (p-aLBGs; Equation~\ref{eq:lsst_z3_rgr_a}) and \lya\ dominant in emission (p-eLBGs; Equation~\ref{eq:lsst_z3_rgr_e}), might be selected from VRO/LSST LBG data dispersed in $(g-r)$ vs.\ $r$ space.

Specifically, for p-aLBGs:
\begin{equation}
\label{eq:lsst_z3_rgr_a} 
(g - r) \ge 0.25 \cdot r - 5.72 + c_{\sigma} \cdot \sigma_e 
\end{equation}

\noindent and for p-eLBGs:

\begin{equation}
\label{eq:lsst_z3_rgr_e} 
(g - r) \le 0.25 \cdot r - 5.72 - c_\sigma \cdot \sigma_a 
\end{equation}

\noindent where $c_{\sigma}$, ${\sigma}_{a}$, and ${\sigma}_{e}$ are the coefficient and respective standard deviations of colour dispersion as described in Section ~\ref{sec:c3_z2_purity}.

These selection criteria provide a useful starting point for the isolation of p-aLBG and p-eLBG populations from LSST photometry.  They will be confirmed and/or refined, and selection criteria in other redshift ranges added -- especially at $z\sim2$ -- once LSST data of sufficient depth has been measured in fields within which \lya\ spectroscopic data are available.

\section{Summary and Conclusions} 

The \lya\ observables from a given galaxy are known to be sensitive to a wide range of galactic physical, spectral, and environmental properties.  Net \lya\ EW in particular has been shown to correlate with, for example, galaxy morphology, rest-frame UV colour, ISM line strengths, gas kinematics and, the large-scale spatial distribution of populations of $z\gtrsim2$ LAEs and LBGs.  Accordingly, the ability to select pure statistical sub-samples of a desired \lya\ spectral type from large photometric datasets facilitates the study of a variety of intrinsic and small- to large-scale environmental galactic properties that are related to \lya, in large numbers and over distance scales for which ancillary multi-wavelength/multi-band photometry and/or spectroscopic information is not usually available \citep{Cooke2009, Cooke2013}.

In this paper we characterise the broadband imaging segregation of a spectroscopic sample of 557 $z\sim2$ LBGs using sub-samples with \lya\ dominant in absorption (aLBGs), and \lya\ dominant in emission (eLBGs), and determine photometric criteria by which relatively pure sub-samples with desired \lya\ spectral properties can be selected using imaging data in as few as three optical broadband filters.  

We draw the following specific conclusions from our study:

\begin{itemize}
\item $z\sim2$ LBGs segregate according to their net \lya\ EW properties in rest-frame UV colour ($U_n-\cal{R}$) and magnitude ($\cal{R}$) space in a manner similar to their $z\sim3$ counterparts in the $(G-\cal{R})$/$\cal{R}$ plane \citep[see Section~\ref{sec:c3_seg} and cf.][]{Cooke2009}.  
\item Using the segregation statistics for our sample of 557 LBGs in the range $1.7<z<2.5$, we determine photometric criteria for the selection of sub-samples of LBGs with \lya\ dominant in absorption (p-aLBGs) and \lya\ dominant in emission (p-eLBGs). These criteria select sub-samples of p-aLBGs and p-eLBGs that are respectively $\gtrsim 97\%$ and $\sim$85\% pure with respect to contamination by galaxies with the opposite spectral type.  The mean net \lya\ EW of the optimised p-aLBG and p-eLBG sub-samples selected from the $z\sim2$ \UGR\ LBGs is $\sim -$8\,\AA\ and $\sim +$8\,\AA, respectively (Section~\ref{sec:c3_z2_purity}).
\item Sub-dividing the $z\sim2$ sample into two redshift bins, we find that the degree of photometric segregation in the range $2.17<z<2.5$ (\lya\ outside the $U_n$ filter) is significantly greater than in the range $1.70<z<2.17$ (\lya\ within the $U_n$ filter).  We attribute this difference to a larger contribution of the \lya\ forest leading to greater dispersion in $(U_n-\cal{R})$ colour at higher redshifts. In the range $2.17<z<2.5$, we select sub-samples of p-aLBGs and p-eLBGs that are $\gtrsim$95\% pure with respect to galaxies of the opposite spectral type, and which segregate in mean net \lya\ EW ($\Delta_{Ly\alpha\ EW} \approx$30 \AA) on the same order as the $z\sim3$ LBGs (Section~\ref{sec:c3_lya}).
\item Using the result of C09 and spectrophotometry of the composite spectra of $z\sim3$ LBGs with different \lya\ spectral type, we calculate photometric criteria by which populations of p-aLBGs and p-eLBGs can be selected from the \ugri\ broadband imaging of the LSST (Section~\ref{sec:lsst}).
\end{itemize}

One motivation for this work is to provide the statistical foundation for application of the result described in Paper\,II in this series (Foran et al.\ (2023b), submitted) wherein we report a relationship between net \lya\ EW and galaxy kinematics, and demonstrate how the photometric segregation described here can be used to predict the kinematic type (and other properties) of large numbers of $z\sim2-3$ LBGs without the need for spectroscopic information.  

More broadly, we propose that this method has strong potential to expand the legacy value of the current generation of deep, wide, optical and near-infrared, large-area and all-sky photometric campaigns such as the Hyper-SuprimeCam Subaru Strategic Program \citep[HSC-SSP:][]{Aihara2018} and the upcoming Vera Rubin Observatory Legacy Survey of Space and Time \citep[LSST:][]{Ivezic2019}, that will exploit the Lyman break technique using 3--5 broadband filters across the rest-frame UV to select hundreds of millions of galaxies in redshift ranges from $z\sim2-6$ across many hundreds to thousands of Mpc.  Optimising the discovery potential of such programs requires new techniques to statistically characterise such huge datasets, and to efficiently select from these the most promising samples for expensive follow-up observations.  The techniques and insights presented here and in Paper II, explore how inexpensive broadband photometric information that is sensitive to the \lya\ properties of LBGs might address this challenge.  This approach also provides a statistical framework within which $z\sim2-3$ LBGs will serve as low-redshift reference samples for the study of galaxy populations at higher redshifts where only selection methods based on \lya\ emission or Lyman break detection can be applied  in large numbers and over large scales \citep{Finkelstein2016}.

Specific applications of this approach might include:

\begin{itemize} 
\item study of the environments of \lya\ absorbers and emitters on small and large scales out to hundreds and thousands of Mpc; 
\item generation of the large samples of \lya\ absorbers and emitters required for three-point correlation function analysis, whereby the geometry, spatial shape, and distribution of the different spectral types might be mapped relative to the filaments and nodes of the cosmic web; 
\item investigation of the origins and character of the Morphology--Density Relation at $z\sim2$ and beyond; 
\item furnishing of the kinematic properties of large numbers of early galaxies of known \lya\ spectral-type to aid halo-matching between observations and cosmological simulations; and 
\item cosmological studies in which tailored samples of $z\sim2-5$ LBGs with varying \lya\ EW are used in combination with Cosmic Microwave Background (CMB) lensing cross-correlation analysis, to infer the time evolution of matter-density fluctuations, and  to carry out compelling tests of horizon-scale General Relativity, neutrino masses and Inflation \citep[e.g.,][]{Wilson2019}.  
\end{itemize}

As a first step toward these goals, we present here photometric criteria by which populations of $z\sim3$ LBGs with \lya\ dominant in absorption, and \lya\ dominant in emission, might be selected from \ugri\ photometric data from the LSST (Section~\ref{sec:lsst}).

\begin{acknowledgement}
None.
\end{acknowledgement}


\bibliography{all_refs}

\appendix

\section{Photometric Uncertainties}
\label{sec:photo_unc}

As part of corrections for photometric incompleteness in their study of the rest-frame UV luminosity function at $z\sim1.9-3.4$, \citet{Reddy2008} applied a Monte Carlo (MC) galaxy population simulation method  to joint photometric and spectroscopic samples of $z\sim2$ \UGR\ LBGs to assess the systematic effects of photometric scatter and the intrinsic variation in colours due to \lya\ line emission and absorption.  These simulations yielded statistical estimates of the photometric uncertainties for the imaging data used to select the $z\sim2$ and $z\sim3$ \UGR\ LBG samples used in this work.  Tables of these uncertainties were supplied (N.\ Reddy, priv.\ comm.) in 0.5 mag bins of $G$ and $\cal{R}$ magnitude and 0.2 mag bins of $(U_n-G)$ and $(G-\cal{R})$ colour for all observed fields in the $z\sim2$ \UGR\ survey.  In the absence of source-by-source photometric errors, we calculated from the MC simulation data indicative photometric uncertainties for the $z\sim2$ and $z\sim3$ LBGs dispersed in $(U_n-\cal{R})$/$\cal{R}$ and $(G-\cal{R})$/$\cal{R}$ colour/magnitude space, respectively.

For the $z\sim2$ LBGs, $\cal{R}$-band and $(G-\cal{R})$ uncertainties for each galaxy were extracted from the MC simulation tables for the relevant field according to their observed $\cal{R}$-band luminosity and $(G-\cal{R})$ colour, and added in quadrature to give calculated estimates of $G$-band uncertainty.  These $G$-band uncertainties were in turn added in quadrature with the tabulated $(U_n-G)$ uncertainties to give an estimate of the $U_n$-band uncertainty for each galaxy.  Finally, $(U_n-\cal{R})$ uncertainties were estimated by subtracting in quadrature the $\cal{R}$-band uncertainty from that of the $U_n$-band.

For $z\sim3$ LBGs  in the three fields where the $z\sim2$ and $z\sim3$ catalogs overlap (i.e., HDF/GOODS-N, Q0933 and Q1422) $\cal{R}$-band and $(G-\cal{R})$ uncertainties for each galaxy were extracted similarly to the $z\sim2$ sample.  For all other $z\sim3$ LBGs, $\cal{R}$-band and $(G-\cal{R})$ uncertainties were estimated by averaging values for the HDF/GOODS-N, Q0933 and Q1422 fields at the relevant luminosity and colour.

\begin{figure} 
\begin{centering} 
\includegraphics[width=\columnwidth]{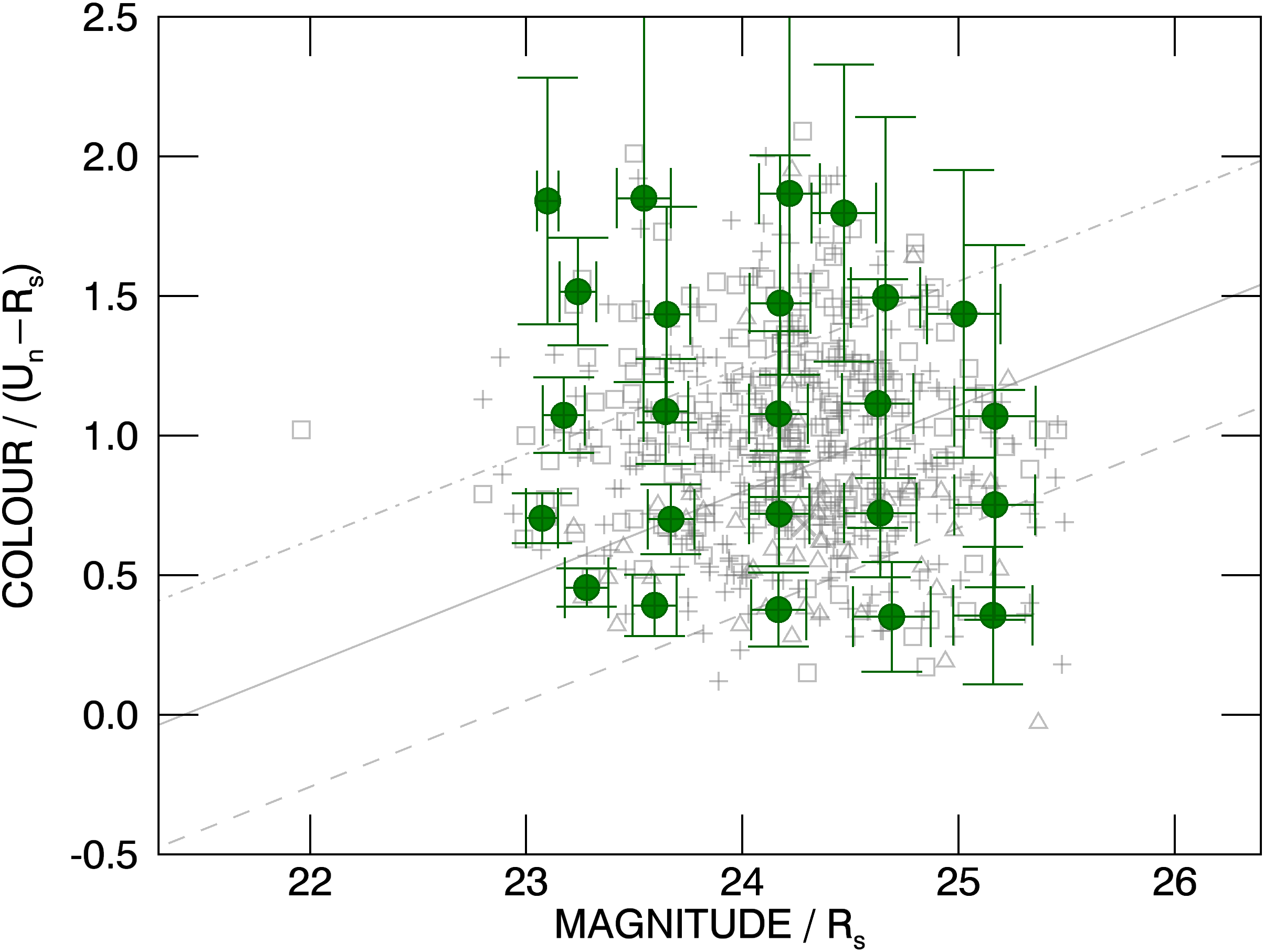}
\includegraphics[width=\columnwidth]{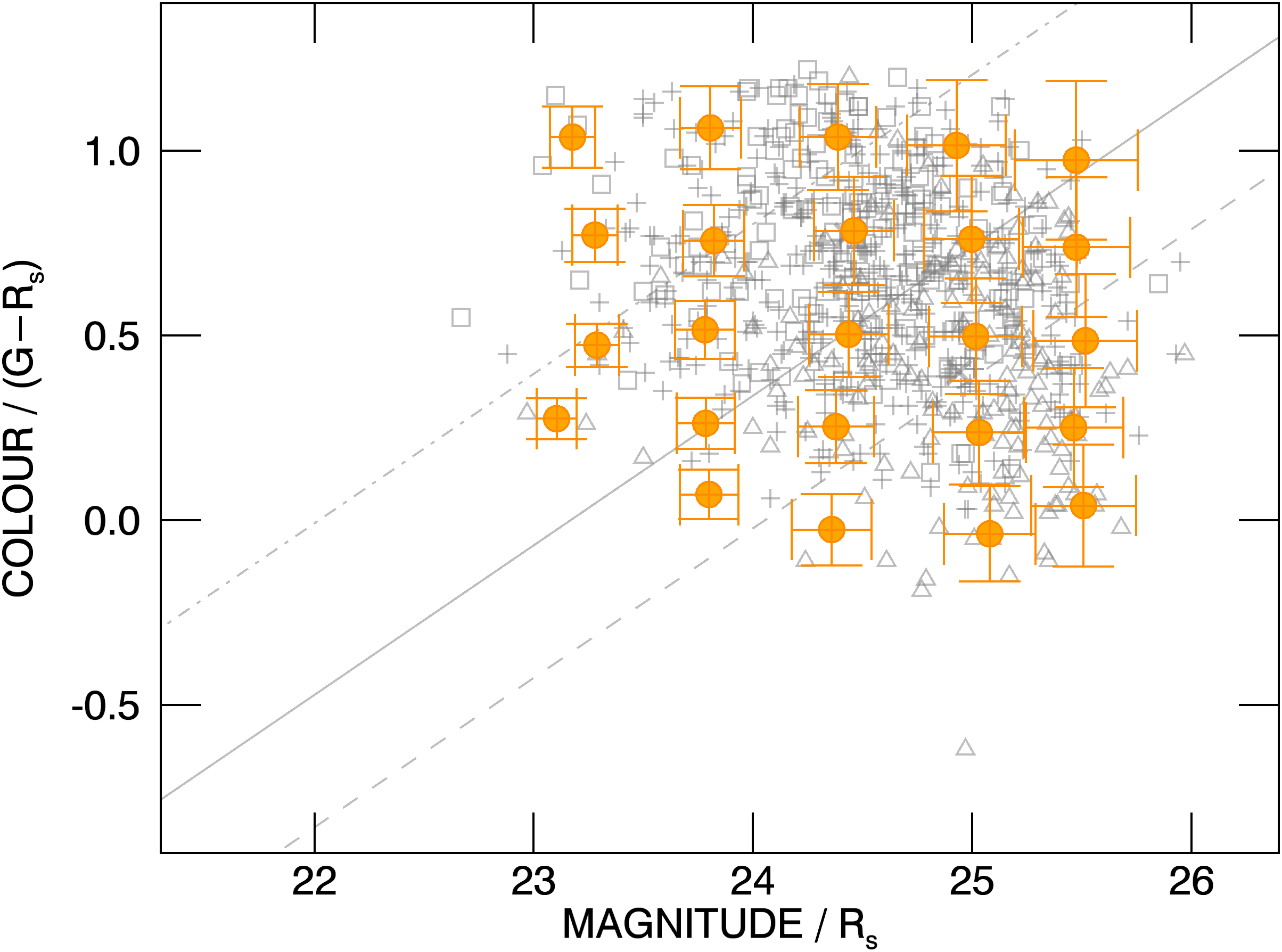}
\caption{Indicative photometric uncertainties for $z\sim2$ (top) and $z\sim3$ (bottom) \UGR\ LBGs dispersed in colour--magnitude space and divided into a $5\times5$ grid on the CMD.  The green and orange symbols indicate the mean colour, magnitude, and associated uncertainties for the galaxies in each grid element.  In each case, the representative symbols are overlaid on their respective full sample (grey symbols).}
\label{fig:c2_lbg_errors}
\end{centering} 
\end{figure}

For the purpose of illustrating the representative photometric uncertainties thus calculated, the $z\sim2$ and $z\sim3$ LBG samples were divided into a $5\times5$ grid on their respective CMDs.  Figure~\ref{fig:c2_lbg_errors} shows the mean colour, magnitude, and associated uncertainties, for the galaxies in each grid element overlaid on the full sample for both redshift ranges.

The estimated typical $\cal{R}$-band uncertainty of $\lesssim 0.2$ up to the $\cal{R}$ $= 25.5$ limit, gives confidence for the use of the $z\sim2$ \UGR\ LBG sample in our analysis.  On the other hand, the estimated uncertainties in $(U_n-\cal{R})$ suggest that $U_n$-band magnitudes fainter than $\sim26.0-26.5$ introduce photometric errors $\gtrsim 0.5$ that are potentially problematic for the colour--magnitude segregation approach investigated here.  Given our need for reliable $(U_n-G)$ and/or $(U_n-\cal{R})$ colours, and in the absence of source-by-source photometric uncertainties, the estimates derived from the MC simulations motivated the decision to limit the $z\sim2$ sample to galaxies with $U_n$-band magnitudes $\leq$ 26.5.

\end{document}